\documentclass[epj, final]{svjour}

\pdfoutput=1

\usepackage{graphicx}
\usepackage{amsmath}
\usepackage{amssymb}
\usepackage{cite}
\usepackage{units}
\usepackage{url}

\DeclareGraphicsExtensions{.pdf}

\begin{document}

\title{The embedding method beyond the single-channel case} 
\subtitle{Two-mode and Hubbard chains} 

\author{Axel Freyn\inst{1,2} \and Gabriel Vasseur\inst{3} 
\and Peter Schmitteckert\inst{4} \and Dietmar Weinmann\inst{3}
\and Gert-Ludwig Ingold\inst{5} \and Rodolfo A.\ Jalabert\inst{3} 
\and Jean-Louis Pichard\inst{1}}

\authorrunning{A.\ Freyn \textit{et al.}}

\institute{
Service de Physique de l'{\'E}tat Condens{\'e} (CNRS URA 2464),
  IRAMIS/SPEC,
  CEA Saclay, 
  91191 Gif-sur-Yvette Cedex, France
\and
Institut N{\'e}el,
  25 avenue des Martyrs,
  BP 166,
  38042 Grenoble Cedex 9, France
\and
  Institut de Physique et Chimie des Mat\'eriaux de Strasbourg,
  UMR 7504 (UdS-CNRS), 23 rue du Loess, BP 43, 
  67034 Strasbourg Cedex 2, France
\and
  Institut f\"ur Nanotechnologie,
  Forschungszentrum Karlsruhe,
  Karlsruhe Institute of Technology, Postfach 3640,
  76021~Karls\-ruhe, Germany
\and 
  Institut f\"ur Physik,
  Universit\"at Augsburg, Universit\"atsstra{\ss}e 1, 
  86135 Augsburg, Germany
}
\date{\today}

\abstract{
We investigate the relationship between persistent currents in multi-channel 
rings containing an embedded scatterer and the conductance through the same 
scatterer attached to leads. The case of two uncoupled channels corresponds to 
a Hubbard chain, for which the one-dimensional embedding method is readily 
generalized. Various tests are carried out to validate this new procedure, and 
the conductance of short one-dimensional Hubbard chains attached to perfect 
leads is computed for different system sizes and interaction strengths. 
In the case of two coupled channels the conductance can be obtained from a 
statistical analysis of the persistent current or by reducing the 
multi-channel scattering problem to several single-channel setups. 
} 

\maketitle
\sloppy

\section{Introduction}

Electronic correlations can influence coherent electronic transport in striking
ways. An example is the Kondo effect in the transport through quantum dots 
\cite{goldhaber98,pustilnik01,hewson_book}. It is well-known that the inclusion
of the many-body effects arising from electron-electron interactions in quantum
transport calculations is extremely difficult. When a coherent and interacting
nanosystem is attached to Fermi liquid reservoirs via leads, the fundamental
problem consists in the matching of the correlated many-body wave-function in 
the nanosystem with the effective one-body wave-functions in the leads that
allow to define the bias voltage as difference between the chemical potentials
of the reservoirs. Even for the simplest nanosystems such a matching procedure
is very complicated to apply \cite{mehta06,doyon07,boulat08}. There is an 
increasing literature where the auxiliary Kohn-Sham wave-functions of density 
functional calculations are used to determine the transport properties of 
realistic systems. However, no rigorous theoretical justification for this 
approach exists at present and approximations in the density functional can 
considerably influence the results for transport \cite{schmitteckert07}. 

At sufficiently low temperatures and when the interactions are restricted to 
a small region in space that is attached to noninteracting leads, a 
Landauer-like formula with an effective interaction-dependent transmission 
\cite{meir92} holds for the linear (dimensionless) conductance, 
\begin{equation}\label{mccond}
g=\text{Tr}\{\tens{t}\tens{t}^\dagger\}\,.
\end{equation}
For leads supporting $N$ open channels, the transmission matrix 
$\tens{t}$ is of dimension $N$. Determining this 
transmission matrix remains however a hard task. 

The recently developed embedding method 
\cite{favand98,sushkov01,molina03,meden03,rejec03,molina04a,molina05}
represents a considerable advance for the single-channel case, $N=1$, 
allowing to 
extract the effective transmission probability $|t|^2$ of the nanosystem at 
the Fermi energy of the reservoirs. This method is based on the fact that the 
persistent current through a ring, made of a scatterer and a non-interacting 
lead and threaded by a magnetic flux $\phi$, depends on the transmission 
probability of the scatterer. At zero temperature the flux 
dependence of the ground-state energy $E_0$ determines the persistent current 
through \begin{equation}\label{def_pers_curr}
J=-\frac{\partial E_0}{\partial\phi}\,.
\end{equation}
In the limit of very large lead length $L_\mathrm{L}$ and for a non-interacting
scatterer, the persistent current and the transmission probability can be 
easily related. For an odd number of spinless fermions in the ring, the flux
dependence of the persistent current in lowest order in $1/L$ as a function of 
the transmission amplitude $t$ at the Fermi wavenumber $k_\text{F}$
is given by \cite{gogolin94}
\begin{equation}
J(\Phi) = -\frac{ev_\mathrm{F}}{\pi L} \frac{\mathrm{arccos}
              \big(|t(k_\mathrm{F})| \cos(\Phi)\big)}
     {\sqrt{1- |t(k_\mathrm{F})|^2\cos^2(\Phi)}} |t(k_\mathrm{F})| \sin(\Phi)\,.
\label{gogolin-pc}
\end{equation}
Here, $L_\mathrm{S}$ denotes the length of the nanosystem, 
$L=L_\mathrm{S}+L_\mathrm{L}$ is the total length of the ring, and 
$\Phi=(e/\hbar)\phi$ the dimensionless flux.
	
A similar expression holds for an even number of fer\-mi\-ons. In particular,
\begin{equation}\label{t-pers_curr}
|t|^2 = \lim_{L\rightarrow \infty}
\left(\frac{J_L(\pi/2)}{J^{0}_L(\pi/2)}\right)^{2}\,,
\end{equation}
where $J^{0}_L(\pi/2)$ is the persistent current of a perfect ring without 
scatterer at flux $\Phi=\pi/2$.

Using the same relationships for the interacting case allows to extract $|t|$ 
from the many-body ground state properties of the ring. It has been shown 
\cite{molina04a} that for an interacting single-channel scattering problem the 
full flux dependence of the persistent current can be described by 
\eqref{gogolin-pc}. Such a check provides a strong evidence for the validity 
of the embedding method. A nanosytem can then be replaced by an effective 
one-body scatterer, provided the electrodes are connected to the interacting 
region by sufficiently long noninteracting leads \cite{molina05, weinmann08}. 

A convenient measure for the persistent current is the phase sensitivity 
$\mathcal{D}= (L/2)|E_{0,\mathrm{P}} - E_{0,\mathrm{A}}|$, where  
$E_{0,\mathrm{P}}$ and $E_{0,\mathrm{A}}$ denote the ground state energies of 
the ring for periodic and anti-periodic boundary conditions, respectively. 
An extrapolation of $\mathcal{D}$ for $L \to \infty$ keeping $L_\mathrm{S}$ 
constant yields $\mathcal{D}_{\infty}$ and the absolute value of the effective 
transmission amplitude of the system at the Fermi energy 
\begin{equation}\label{t-sensitivity}
|t| = \sin\left(\frac{\pi}{2}
                \frac{\mathcal{D}_{\infty}}{\mathcal{D}_{\infty}^0}
          \right)\,.
\end{equation}
Here, $\mathcal{D}_{\infty}^0=\pi\hbar v_\mathrm{F}/4$, with the Fermi velocity 
$v_\mathrm{F}$, is the phase sensitivity of a ring with perfect transmission 
$t=1$ through the scatterer. 

Another quantity that can be used to characterize the flux dependence of 
the persistent current is the curvature
\begin{equation}
\mathcal{C}=L\frac{e^2}{\hbar^2}
          \left.\frac{\partial^2E}{\partial\Phi^2}\right|_{\Phi=0} 
\end{equation}   
of the many-body ground state energy as a function of the flux. Because of a 
level crossing occurring at zero flux, the curvature diverges for the case of 
an even number of particles in a clean ring.  For an odd number of particles, 
the curvature is well-behaved. Its long-lead limit $L\to\infty$ 
\begin{equation}\label{t-curvature}
\mathcal{C}_{\infty}=\frac{e^2 v_\mathrm{F}}{\pi\hbar}
                    \frac{|t|\arccos(|t|)}{\sqrt{1-|t|^2}}
\end{equation}
can be obtained from \eqref{gogolin-pc}, and depends only on the absolute 
value of the transmission amplitude \cite{vasseur_thesis}. 
The inverse of \eqref{t-curvature} is unique, but in general 
$\vert t\vert$ has to be determined numerically from 
$\mathcal{C}_\infty$.

It is important to stress that the previous relations between the persistent 
current and the transmission probability at the Fermi energy have only been 
established for the case of spinless fermions with strictly one-dimensional 
leads. According to \eqref{mccond}, the conductance in the case of more than 
one channel is given as the sum over all transmission probabilities between 
pairs of channels. Then, the value of the persistent current alone is in 
general no longer sufficient to determine the conductance.

In the present paper we generalize the embedding meth\-od to fermions with spin 
and multi-channel situations. After a discussion of persistent currents in 
rings embedding a multi-channel scatterer (Sec.~\ref{sec:rcm}), we address the 
one-dimensional Hubbard chain attached to perfect leads 
(Sec.~\ref{sec:Hubbard}). In this latter case, because of the spin rotation 
symmetry it is straightforward to extend the embedding method developed for the
single-channel case in order to include spin. We discuss some technical aspects
of the evaluation of the conductance, develop an improved embedding method 
based on damped boundary conditions, and present numerical results for short 
chains. 

Two paths are presented in order to treat the case of two coupled channels. 
The first one is a statistical study of the persistent currents with random 
channel mixing (Sec.~\ref{section_pcftcwrcm}). The second one is a reduction 
of the multi-channel problem to several single-channel problems, allowing to 
infer the conductance from the corresponding single-channel transmission 
coefficients that can be determined using the standard procedure of the 
embedding method (Sec.~\ref{sec:reduction} and Appendix~\ref{appendix_eocp}). 
This second path is used in order to argue that channel mixing is absent in 
the case of Hubbard chains with spin-rotation symmetry. We have relegated to 
the appendices some of the lengthy formulations and alternative proofs. 

\section{Persistent currents in multi-channel rings embedding a scatterer}
\label{sec:rcm}

In this section we address the relationship between the transmissions of a 
multi-channel non-interacting scatterer and the persistent current in rings 
embedding the scatterer. In particular, we underline the difficulties in
going from one to two channels. 

\subsection{\boldmath{$N$}-channel scattering problem}

The ideal quasi-one-dimensional leads are characterized by translational 
invariance along the wire axis and a position-independent lateral 
confinement, allowing to separate the one-body Hamiltonian into a 
longitudinal and a transverse part.
The solutions of the longitudinal part are plane waves and the eigenstates 
$\phi_n(y)$ of the transverse part give rise to the conduction channels and 
energy offsets $\varepsilon_n$. It is convenient to describe the states 
in the lead at a given energy $\varepsilon$ as a superposition of 
flux-normalized one-dimensional plane waves as
\begin{equation}
\label{psin}
\Psi(x,y)= \sum_{n=1}^{N}\frac{1}{\sqrt{k_n}}\left[A_n\exp{(\mathrm{i}k_nx)}
      +B_n\exp{(-\mathrm{i}k_nx)}\right]\phi_n(y)\, ,
\end{equation}
where $k_n$ denotes the wave vector in the $n$th channel. In general, the 
channel index $n$ also accounts for the electron spin as in the case 
of the Hubbard chain discussed in Section~\ref{sec:Hubbard}.

\begin{figure}
\centerline{\includegraphics[width=0.4\columnwidth]{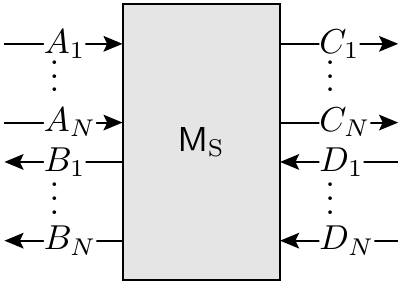}}
\caption{\label{fig:scatter}Sketch of the transfer-matrix description of 
scattering in multi-channel wires.}
\end{figure}

As represented in Figure~\ref{fig:scatter}, the transfer matrix 
$\tens{M}_\text{S}$ relates the amplitudes $A_n$ and $B_n$
of $\Psi_\mathrm{I}$ on the left-hand side to the corresponding amplitudes
$C_n$ and $D_n$ of $\Psi_\mathrm{II}$ on the right-hand side of the scatterer 
according to
\begin{equation}
\label{eq:cd_ms_ab}
\begin{pmatrix} C_1 \\ \vdots \\ C_N \\ D_1 \\ \vdots \\ D_N \end{pmatrix}
=\tens{M}_\mathrm{S}
\begin{pmatrix} A_1 \\ \vdots \\ A_N \\ B_1 \\ \vdots \\ B_N \end{pmatrix}\,.
\end{equation} 

\begin{figure}
\centerline{\includegraphics[width=0.8\columnwidth]{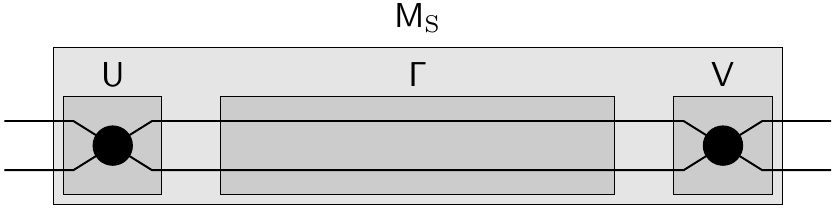}}
\caption{\label{fig:mixing}A general transfer matrix
$\tens{M}_\text{S}$ satisfying current conservation and time-reversal 
symmetry can be expressed as the product \eqref{MS} of three matrices.
$\tens{U}$ and $\tens{V}$ relate the eigenmodes in the
leads to those in the system and $\tens{\Gamma}$ describes the
transmission and reflection properties of the system.} 
\end{figure}

Assuming current conservation and time reversal symmetry, it is possible to 
express the transfer matrix of an $N$-channel scatterer in the polar 
decomposition
\begin{equation}\label{MS}
\tens{M}_\mathrm{S} =\tens{U\Gamma V}
\end{equation}
as a product of three matrices, where  
\begin{equation}\label{uvlambda}
  \begin{aligned}
  \tens{U} &= \begin{pmatrix} \tens{u}& 0 \\ 
                                     0 & \tens{u}^* \end{pmatrix},\\
  \tens{V} &= \begin{pmatrix} \tens{v} & 0 \\ 
                                     0 & \tens{v}^* \end{pmatrix},\\
  \tens{\Gamma} &= \begin{pmatrix} \sqrt{\tens{1}+\tens{\lambda}} & \sqrt{\tens{\lambda}} \\
             \sqrt{\tens{\lambda}} & \sqrt{\tens{1}+\tens{\lambda}} \end{pmatrix}\,.
  \end{aligned}
\end{equation}
$\tens{u}$ and $\tens{v}$ are complex unitary $N\times N$-matrices, and 
$\tens{\lambda}$ is a real, diagonal $N\times N$-matrix \cite{mello91}. 
The matrix $\tens{\lambda}$ contains all transmission probabilities 
$|t_a|^2=1/(1+\lambda_a)$ for the different eigenmodes of the scatterer, 
where $\lambda_a$ is the $a$\textsuperscript{th} entry of $\tens{\lambda}$. 
Therefore, $\tens{\lambda}$ determines the conductance of the scatterer through 
$g=\sum_{a=1}^N 1/(1+\lambda_a)$. The matrices $\tens{u}$ and 
$\tens{v}$ describe the way these eigenmodes are connected to the 
different incoming and outgoing channels. In general, a mixing of channels as 
illustrated in Figure~\ref{fig:mixing} will occur. The main interest of the polar
decomposition consists in the very different r{\^o}le played by the radial 
($\tens{\lambda}$) and angular parameters ($\tens{u}, \tens{v}$), 
allowing the development of random matrix theories to describe the quantum 
transport through an arbitrary scatterer \cite{been_97,jal_95}.

An $N$-channel ring consisting of a scatterer described by 
$\tens{M}_\mathrm{S}$  and an ideal lead can be characterized 
by the total transfer matrix of the ring 
$\tens{M}_\mathrm{S}\tens{M}_\mathrm{L}$, where $\tens{M}_\mathrm{L}$ 
is the diagonal transfer matrix of the lead with entries
\begin{equation}\label{ml}
  \begin{aligned}
  \left(\tens{M}_\mathrm{L}\right)_{nn} &=  
       \exp{(\mathrm{i}\Phi+\mathrm{i}k_nL_\mathrm{L})} ,\\
   \left(\tens{M}_\mathrm{L}\right)_{n+N,n+N} &= 
        \exp{(\mathrm{i}\Phi-\mathrm{i}k_{n}L_\mathrm{L})}\, ,
  \end{aligned}
\end{equation}
describing right- and left-moving electrons ($1\le n\le N$).

The one-particle eigenstates $\psi^{(p)}$ of the ring are defined by
the amplitudes $A_n^{(p)}$ and $B_n^{(p)}$ through the matrix condition
\begin{equation}\label{ring-quant_cond}
(\tens{1}-\tens{M}_\mathrm{S}\tens{M}_\mathrm{L})
\begin{pmatrix}\vec{A}^{(p)}\\
\vec{B}^{(p)}\end{pmatrix}=0\,.
\end{equation}
The corresponding eigenenergies $\varepsilon^{(p)}$ satisfy 
\begin{equation}\label{ring-transfer}
\text{det}(\tens{1}-\tens{M}_\mathrm{S}\tens{M}_\mathrm{L})=0\, ,
\end{equation}
where $\tens{1}$ denotes the $2N\times2N$ identity matrix.
The contribution $j^{(p)}$ to the persistent current of the ring follows
from the flux dependence of $\varepsilon^{(p)}$ through
Equation~\eqref{ring-transfer}. Alternatively, it can also be obtained from the 
expectation value of the current operator in the leads  in the limit of 
very large $L_\mathrm{L}$, 
\begin{equation}\label{persistent_current}
j^{(p)}=\frac{e\hbar}{m} \sum_{n=1}^{N}
\left(|A_n^{(p)}|^2-|B_n^{(p)}|^2\right)\, .
\end{equation}  

Summing the contributions of all occupied states $p$ we are lead to the 
persistent current through the ring.

\subsection{One-channel scattering problem}

For energies  $\varepsilon$ between the two lowest transverse energies
$\varepsilon_1$ and $\varepsilon_2$ only one channel propagates
and the problem is particularly simple \cite{molina04a} since
$\tens{u}$ and $\tens{v}$ are pure phases given,
respectively, by $\exp[\mathrm{i}(\alpha \pm \zeta)/2]$ in terms of
the angles $\alpha$ and $\zeta$. The quantization condition
\eqref{ring-transfer} can be written as
\begin{equation}\label{qc1d}
\cos(\Phi)=\frac{1}{\sin(\theta)\sin(\varphi)} \cos(kL +\delta \alpha)\, .
\end{equation}   

Following the standard parametrization of a $2\times 2$ transfer matrix
\cite{molina04a}, we have defined the angles $\theta$ and $\varphi$ 
through
\begin{equation}\label{defang}
  \begin{aligned}
   \sin(\theta) &= \sqrt{\frac{1+\lambda \cos^2(\zeta)}{1+\lambda}} ,\\
   \sin(\varphi) &= \frac{1}{\sqrt{1+\lambda\cos^2(\zeta)}}\, .
  \end{aligned}
\end{equation}
In this way $t=\exp(\mathrm{i}\alpha)\sin(\theta)\sin(\varphi)$ 
is the transmission amplitude and
$\delta \alpha=\alpha-kL_\mathrm{S}$ is the scattering phase-shift.
We remark in passing that the expressions simplify in the particular 
case with right-left symmetry, where $\varphi=\zeta=\pi/2$.

Taking the flux derivative for the $p$-th solution of Equation~\eqref{qc1d} 
and following the line of reasoning presented in the appendix of 
\cite{molina04a} we obtain the contribution $j^{(p)}$ which
after summation over the occupied states leads to \eqref{gogolin-pc}. 
For the particular case of $\Phi=\pi/2$ we have \cite{molina04a}
\begin{equation}\label{jp}
j^{(p)}(\pi/2)= (-1)^p \ \frac{e\hbar}{L m} \ k^{(p)} |t(k^{(p)})| \, .
\end{equation}  

The alternative route through the eigenvectors is also simple in the 
$N=1$ case since Equation~\eqref{persistent_current} becomes
\begin{equation}\label{persistent_current_Neq1}
j^{(p)}=\frac{e \hbar}{mL} \ k^{(p)} \left(\frac{1-|\chi^{(p)}|^2}
{1+|\chi^{(p)}|^2}\right) \, , 
\end{equation}  
with
\begin{equation}\label{chi_sq_Neq1}
\begin{aligned}
|\chi^{(p)}|^2 & =\left|\frac{B_1^{(p)}}{A_1^{(p)}}\right|^2=
\frac{1}{1-\sin^2(\theta)\sin^2(\varphi)} \\
& \times \big[1+\sin^2(\theta)\sin^2(\varphi)-
2\sin^2(\theta)\sin^2(\varphi)\cos^2(\Phi) - \\
&\hphantom{ \times \big[ } 2\sin(\theta)\sin(\varphi)\sin(\Phi)\\
&\hphantom{ \times \big[ } \times \sin\left[\pm 
\mathrm{arccos}{(\sin(\theta)\sin(\varphi)\cos(\Phi))}\right]
\big],
\end{aligned} 
\end{equation}
where the $\pm$ refers to the parity of the eigenstate label $p$. Using
\begin{equation}\label{chi_sq_Neq1_phieqphio2}
|\chi^{(p)}|^2_{\Phi=\pi/2} =\frac{(1 \mp \sin(\theta)\sin(\varphi))^2} 
{1-\sin^2(\theta)\sin^2(\varphi)}
\end{equation}
in Equation~\eqref{persistent_current_Neq1} we are lead to the contribution 
\eqref{jp} to the persistent current.

\subsection{Two-channel scattering problem}

For energies $\varepsilon$ satisfying 
$\varepsilon_2 < \varepsilon < \varepsilon_3$, we have two propagating 
channels and the matrices $\tens{u}$, $\tens{v}$ 
and $\tens{\lambda}$ appearing in \eqref{uvlambda} can be 
parametrized as
\begin{equation}\label{u1v1lambda1}
  \begin{aligned}
  \tens{u} 
         &= \begin{pmatrix} \cos(\varphi)\mathrm{e}^{\mathrm{i}\alpha} 
                      & \sin(\varphi)\mathrm{e}^{\mathrm{i}(\alpha+\beta)} \\
                       -\sin(\varphi)\mathrm{e}^{\mathrm{i}\gamma} 
                      & \cos(\varphi)\mathrm{e}^{\mathrm{i}(\gamma+\beta)} 
         \end{pmatrix},\\
  \tens{v} 
         &= \begin{pmatrix}\mathrm{i}\cos(\psi)\mathrm{e}^{\mathrm{i}\epsilon} 
         & -\mathrm{i}\sin(\psi)\mathrm{e}^{\mathrm{i}\theta} \\
           -\mathrm{i}\sin(\psi)\mathrm{e}^{\mathrm{i}(\epsilon+\eta)} 
         & -\mathrm{i}\cos(\psi)\mathrm{e}^{\mathrm{i}(\theta+\eta)}
         \end{pmatrix},\\             
  \tens{\lambda} &= \begin{pmatrix} \lambda_1 & 0 \\
                                     0 &  \lambda_2 \end{pmatrix},
  \end{aligned}
\end{equation}
where $\lambda_i=r_i^2/t_i^2$, with $0\le t_i, r_i\le1$ and $t_i^2+r_i^2=1$.
The mixing angle $\varphi$ introduced here should not be confused with the
angle $\varphi$ introduced above for the one-channel case.

The quantization condition \eqref{ring-transfer} can now be written as
\begin{equation}\label{ring-transfer_2}
2 \cos^2(\Phi)-\text{Tr}\{\tens{M}\}\cos(\Phi)+F(\tens{M})-1=0 \, ,
\end{equation} 
where $\tens{M}=\tens{M}_\mathrm{S}\tens{M}_\mathrm{L}(\Phi=0)$
and $F(\tens{M})$ is given by the $2\times 2$ sub-determinants of 
$\tens{M}$ as
\begin{equation}\label{F_subdeterminants}
F(\tens{M}) = \text{Re} \sum_{j=2}^{4}  \begin{vmatrix}
M_{11}   &  M_{1j}  \\
M_{j1}   &  M_{jj}
\end{vmatrix}\ .
\end{equation}

Assuming a parabolic dispersion relation in the leads, $k_1$ and $k_2$
are related by $\hbar^2(k_1^2-k_2^2)/2m=\varepsilon_2-\varepsilon_1=
\Delta \varepsilon$. Defining $k=(k_1+k_2)/2$ we have 
$k_1-k_2=m \Delta \varepsilon/\hbar^2k$ and the quantization condition 
\eqref{ring-transfer_2} at $\Phi=\pi/2$ can be expressed as
\begin{equation}\label{ring-transfer_3}
F(k,L_\mathrm{L},t_1,t_2,\varphi,\alpha,\beta,\gamma,\psi,
\epsilon,\eta,\theta)=1
\, .
\end{equation}
The allowed $k^{(p)}$ define the eigenstates of the ring, with associated 
persistent currents
\begin{equation}\label{per_curr_2}
j^{(p)}(\pi/2)=\frac{e\hbar k^{(p)}}{m}
\left.\left(\frac{\text{Tr}\{\tens{M}\}}{\text{d}F/\text{d}k}
\right)\right|_{k=k^{(p)}} \, .
\end{equation} 

In Appendix~\ref{appendix_ftchp} we give the expressions of 
$\text{Tr}\{\tens{M}\}$ and $F(\tens{M})$ in terms 
of $k_1 \pm k_2$, $L_\mathrm{L}$, $t_{1,2}$, and the angles 
characterizing the matrices 
$\tens{u}$ and $\tens{v}$. The complexity of
such expressions forces us to introduce several simplifications in
our problem. The most radical one is to take the two channels as
spin-degenerate modes. This approximation is treated in the next
section, where we analyze the Hubbard chain and show that most
of the complexities of the two-channel problem are not present in
this case. In Section~\ref{section_pcftcwrcm} we consider a simplified
two-channel problem and introduce statistical concepts to extract
the transmission coefficients.

\section{Conductance of a Hubbard chain}
\label{sec:Hubbard}

In this section, we consider electron transport in a chain with Hubbard-like 
electron-electron interactions in a finite segment of the chain. This is a 
particular case of a two-channel chain where the spin degree of freedom of the 
electrons gives rise to the two channels. We will show that the embedding 
method can be applied in this special case in a quite straightforward manner 
and precise numerical calculations can be carried out.

\subsection{Model, conductance and spin-rotation symmetry}

The Hamiltonian of the whole system reads 
\begin{equation}\label{eq:hamiltonian}
H = H_\mathrm{K} + H_\mathrm{U} \, , 
\end{equation}
where 
\begin{equation}
H_\mathrm{K} = -\sum_i\sum_{\sigma=\uparrow,\downarrow} 
( c_{i,\sigma}^\dagger c_{i+1,\sigma}^{\phantom{\dagger}} + \text{h.c.})
\end{equation}
is the homogeneous kinetic energy part describing electrons in an ideal 1D 
chain. Here, $c_{i,\sigma}$ annihilates an electron with spin $\sigma$ on site 
$i$, and we define the energy scale by setting the hopping amplitude equal to
one. The interacting region situated on sites 1 to $L_\mathrm{S}$ is 
distinguished from the rest of the chain solely by the presence of the on-site 
Hubbard interaction 
\begin{equation}
H_\mathrm{U} = U \sum_{i=1}^{L_\mathrm{S}} (\hat{n}_{i,\uparrow} - 1/2) 
                                         (\hat{n}_{i,\downarrow}-1/2)
\end{equation}
with strength $U$. Here, 
$\hat{n}_{i,\sigma}=c_{i,\sigma}^\dagger c_{i,\sigma}^{\phantom{\dagger}}$. Thus, 
our interacting system is a chain of $L_\mathrm{S}$ sites on which a spin-up 
electron interacts with a spin-down electron on the same site.

We fix the Fermi energy in the leads to the center of the band, corresponding 
to half-filling. The interaction term $H_\mathrm{U}$ contains a background 
potential which renders the system particle-hole symmetric and ensures 
half-filling in the interacting region independent of the interaction strength.
    
The two channels in our problem are defined by the two possible orientations 
of the electron spin $\sigma =\{\uparrow,\downarrow\}$. In the absence of a 
magnetic field, the corresponding states in the leads are degenerate.
  
If we are allowed to describe the many-body scattering through the interacting 
segment of the chain by an effective single-electron scattering situation, the 
conductance is given by the sum over all effective transmission probabilities 
as
\begin{equation}\label{conductance-all}
g=\sum_{\sigma,\sigma'}\left|t_{\sigma\sigma'}\right|^2 \,,
\end{equation}
where $t_{\sigma,\sigma'}$ is the transmission amplitude from the incoming 
channel of spin $\sigma$ into the outgoing channel characterized by spin 
$\sigma'$.

We show in the sequel that spin-rotation symmetry imposes serious limitations 
on the transmission coefficients. Since the Hamiltonian is invariant under a 
rotation of the spin basis $\{|{\uparrow}\rangle,|{\downarrow}\rangle\}$, there
can be no preferential spin orientation. As a consequence, the transmission 
matrix must have the same symmetry. Invariance of the transmission matrix 
$\tens{t}$ under arbitrary spin rotations excludes transmissions 
accompanied by a spin flip so that
\begin{equation}\label{tupdown0}
t_{\uparrow\downarrow} = t_{\downarrow\uparrow} = 0\, .
\end{equation}
Otherwise the transmission matrix could be diagonalized by a spin rotation 
thereby violating the spin-rotation symmetry. Furthermore, the transmissions 
with spin conservation have to be equal
\begin{equation}
\label{tdiag}
t_{\uparrow\uparrow} = t_{\downarrow\downarrow} = t\, .
\end{equation}

An alternative way of demonstrating the conditions \eqref{tupdown0} and
\eqref{tdiag} is to use the procedure of the reduction to the single-channel
scattering. We develop this second path in Appendix~\ref{sec:scsappl}.

\subsection{Numerical results for the conductance of Hubbard chains}

As discussed in the previous section, the effective one-body scattering
describing the transmission through a Hubbard chain is characterized by spin
conservation and the two spin channels are not mixed by the effective one-body
scatterer. The transmission through the effective scatterer is therefore
characterized by a single parameter, the spin-independent transmission
amplitude $t$. The conductance \eqref{conductance-all} then simplifies to
\begin{equation}
g=2\left|t\right|^2\, . 
\end{equation}

This expression holds for arbitrary interaction strength in our Hubbard chain. 
The spin appears as a factor of two, as in the case of general one-dimensional 
Fermi liquid systems with spin \cite{rejec03}. 

The relation between persistent current and effective one-body transmission
amplitude which forms the basis of the standard embedding method for the 
single-channel case can now be generalized to two channels without mixing. The 
persistent current is then given by the sum of the contributions from the two 
independent channels. If in addition to the spin-rotation symmetry of the 
Hamiltonian, we choose a spin-independent filling of the two subsystems with 
$N_\uparrow=N_\downarrow$, the two contributions are equal such that the 
two-channel persistent current is twice the single-channel value. As a 
consequence, \eqref{t-sensitivity} is replaced by
\begin{equation}\label{t-sensitivity-2}
|t| = \sin\left(\frac{\pi}{4}
                \frac{\mathcal{D}_{\infty}^{(2)}}{\mathcal{D}_{\infty}^0} 
          \right)\, ,
\end{equation}
where $\mathcal{D}_\infty^{(2)}$ is the numerically obtained two-channel phase 
sensitivity in the limit of a very large ring. When the ring size is not much 
larger than the Hubbard chain, the flux dependence of the current may change 
considerably and reach a very different functional form for 
$L_\mathrm{L}\ll L_\mathrm{S}$ \cite{waintal08}. 

We use the Density Matrix Renormalization Group algorithm (DMRG)
\cite{white92,DMRG_book,schmitteckert_thesis} adapted to the Hubbard model to calculate 
$\mathcal{D}^{(2)}$ for a two-channel ring with a Hubbard segment, taking fully 
into account the electronic correlations. These many-body ground state 
properties are calculated for half-filled rings with increasing sizes $L$ 
corresponding to odd numbers $N_\uparrow=N_\downarrow=L/2$ of electrons for each 
spin channel. We keep up to 6000 states in the DMRG iterations, and perform 
fits to extrapolate the size-dependence of $\mathcal{D}^{(2)}$ to 
$D_{\infty}^{(2)}$ \cite{vasseur_thesis}.

\begin{figure}[t]
\centerline{\includegraphics[width=0.9\columnwidth]{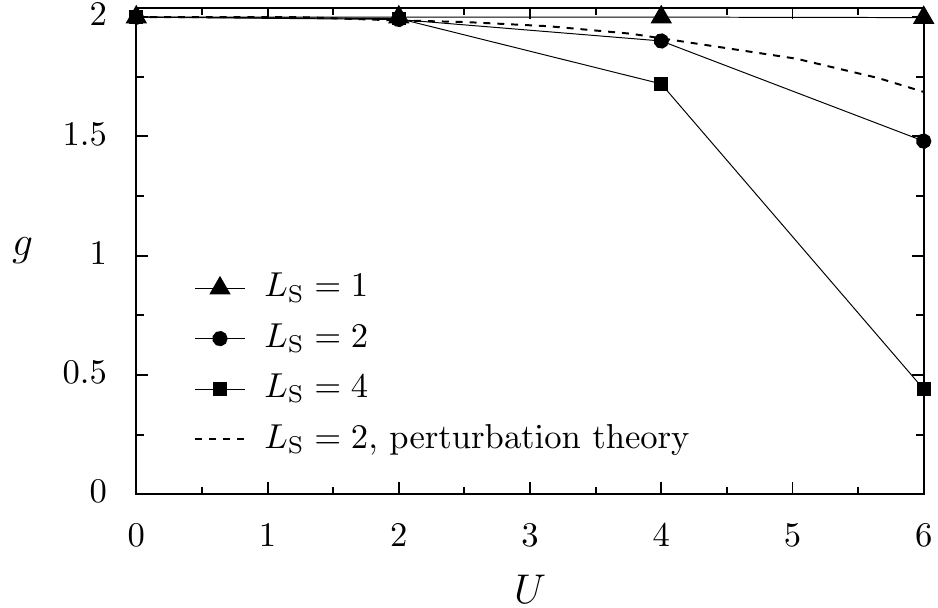}}
\caption{\label{fig:g-hubbard}
Conductance of short Hubbard chains as a function of the interaction strength.
The solid lines are guide to the eyes.
The dashed curve shows the perturbative result produced using the theory of 
Reference~\cite{oguri01}.}
\end{figure}

The numerical results for the conductance of short Hubbard chains using 
\eqref{t-sensitivity-2} are presented in Figure~\ref{fig:g-hubbard}. For even 
$L_\mathrm{S}$ we obtain a suppression of the conductance by the interaction 
that becomes more pronounced for longer Hubbard chains. This can be interpreted 
as a precursor of the Mott-Hubbard transition. For weak interaction strengths, 
our numerical results presented in Figure~\ref{fig:g-hubbard} are in agreement 
with the results of the perturbative approach of Reference~\cite{oguri01}. 
However, the perturbative approach does not yield the significant 
$L_\mathrm{S}$-dependence that we obtain at strong interaction. Nevertheless, a 
strong interaction-induced reduction of the conductance has been obtained for a 
Hubbard chain of length $L_\mathrm{S}$ with reduced coupling to the reservoirs 
\cite{oguri06}.

For odd values of $L_\mathrm{S}$ it has been predicted from perturbative 
arguments \cite{oguri99,oguri01} and a numerical renormalization group (NRG) 
study \cite{oguri05} that the transmission should be perfect. As shown in 
Figure~\ref{fig:g-hubbard}, this perfect conductance $g=2$ is found in the case 
of a single Hubbard impurity $L_\mathrm{S}=1$. For other odd numbers of 
interacting Hubbard sites, our results at strong interaction show small 
deviations from $g=2$ (see Table~\ref{table-Ls3} for the $L_\mathrm{S}=3$ case).
These deviations are of the order of the uncertainty of the extrapolation 
procedure included in the embedding method. It is then important to develop 
complementary approaches to test, and eventually improve, these values. This 
task is undertaken in the next two subsections.

\subsection{Numerical limitations of the standard embedding method}

As mentioned in the introduction, the curvature yields an alternative route for 
extracting the absolute value of the transmission amplitude. In the two-channel 
case, Equation~\eqref{t-curvature} becomes
\begin{equation}\label{t-curvature-2}
\mathcal{C}_{\infty}^{(2)}=\frac{2e^2 v_\mathrm{F}}{\pi\hbar}
                    \frac{|t|\arccos(|t|)}{\sqrt{1-|t|^2}}\,.
\end{equation}

In order to verify that the many-body two-channel scatterer modelled by a chain
with Hubbard-like electron-electron interactions can indeed be replaced by the
effective one-body scatterer without channel mixing described in the previous
section, we compute the transmission from the curvature
$\mathcal{C}_{\infty}^{(2)}$ \cite{schmitteckert04} and compare it with the one
obtained from the phase sensitivity $\mathcal{D}_{\infty}^{(2)}$. Since the
latter is a measure for the integral of the persistent current over the flux
interval $0\le\Phi\le\pi$, and the former corresponds to the slope of the
increase of the persistent current as a function of flux at $\Phi=0$,
$\mathcal{D}_{\infty}^{(2)}$ and $\mathcal{C}_{\infty}^{(2)}$ characterize very
different aspects of the persistent current and its dependence on the flux. The
two quantities are related to the transmission of an effective one-body
scatterer by the relations \eqref{t-sensitivity-2} and \eqref{t-curvature-2}.
If the many-body scatterer can be characterized by an effective one-body
scatterer, then the flux dependence of the persistent current in the correlated
system has to agree with the one of a ring with an effective one-body
scatterer, and the two alternative ways of extracting $|t|$ should be
equivalent. This is the case for even $L_\mathrm{S}$, where the results
obtained from \eqref{t-curvature-2} are very close to the ones resulting from
the phase sensitivity shown in Figure~\ref{fig:g-hubbard}. The strongest
deviations are of the order of 2~\%, which is smaller than the symbol size. For
odd $L_\mathrm{S}$, the results of Table~\ref{table-Ls3} show a difference
between the two methods for large values of the interaction strength. However,
these differences are of the order of the deviation from the expected result of
perfect transmission. Therefore, we can conclude that the two methods agree
within their precision. The consistency of the values of $|t|^2$ extracted from
$\mathcal{D}_{\infty}^{(2)}$ and $\mathcal{C}_{\infty}^{(2)}$ can then be
considered as strong evidence for the validity of the embedding method. 

\begin{table}
\caption{Values of the effective transmission probability $|t|^2$ for a Hubbard
chain of length $L_\mathrm{S}=3$ obtained from the embedding method using the 
phase sensitivity (first line) and the curvature (second line), for different 
values of the interaction strength $U$. The last line is the result of a 
two-parameter fit to the flux dependence obtained using damped boundary 
conditions.}
\label{table-Ls3}
\begin{tabular}{ccccc}\hline\noalign{\smallskip}
          & $U=0$ & $U=2$ & $U=4$ & $U=6$  \\ 
\noalign{\smallskip}\hline\noalign{\smallskip}
from $\mathcal{D}$ & 1.00 & 1.00 & 0.98 & 0.75\\
from $\mathcal{C}$ & 1.00 & 0.99 & 0.95 & 0.65\\
damped boundary    &      & 0.9996 & 0.996 & 0.968\\
\noalign{\smallskip}\hline
\end{tabular}
\end{table}

\begin{figure}[t]
\centerline{\includegraphics[width=0.9\columnwidth]{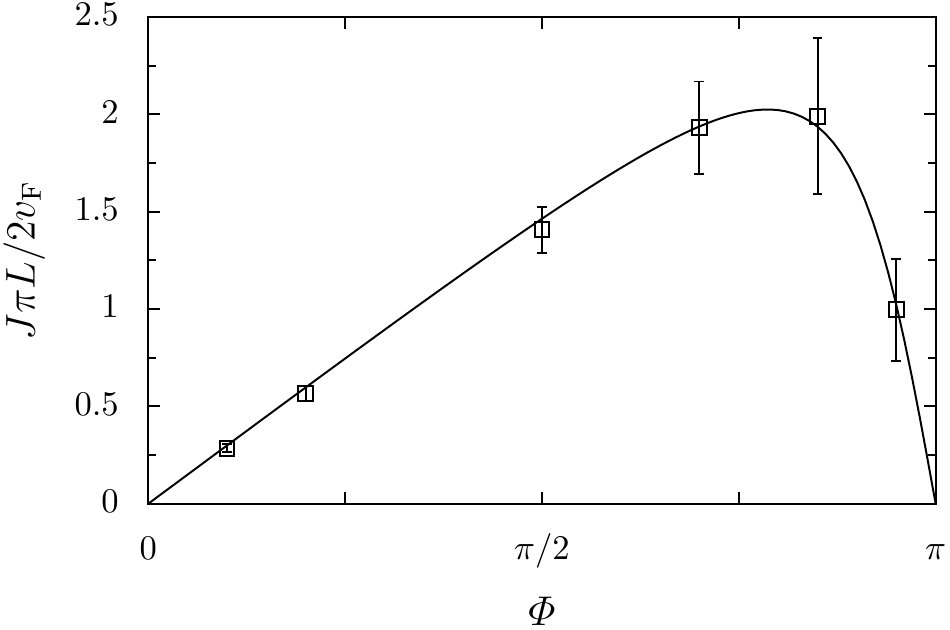}}
\caption{\label{fig:pc-u6}
Persistent currents extrapolated to infinite lead length as a function of the 
magnetic flux for a ring containing a Hubbard chain with $L_\mathrm{S}=3$ and 
$U=6$ (squares). The error bars indicate the difference between a linear and a 
parabolic extrapolation of the available data. The line is a fit of 
\eqref{gogolin-pc} to the data points yielding $|t|^2\approx 0.87$.}
\end{figure}
For the most critical case of Table~\ref{table-Ls3} ($U=6$) we have performed 
the numerically very demanding calculation of the persistent currents for 
different values of the flux. The numerical data extrapolated to infinite lead 
length are shown in Figure~\ref{fig:pc-u6}. The uncertainty of the 
extrapolations indicated by the errorbars remains unfortunately rather 
important. The one-parameter fit of \eqref{gogolin-pc} (with an additional 
factor of 2 for the spin) to the data points yields $|t|^2\approx 0.87$, much 
closer to the expected perfect transmission than the values extracted from 
$\mathcal{D}_{\infty}^{(2)}$ and $\mathcal{C}_{\infty}^{(2)}$ shown in 
Table~\ref{table-Ls3}.

In comparison with the single-channel case, the correlated two-channel
scattering problem requires considerably higher numerical efforts at equal ring
length. As a consequence, we had to limit our computations to total ring sizes
$L\le 30$. As discussed above, this is not generally a problem, but it affects 
the quality of the extrapolations to infinite ring size for situations close 
to conductance resonances \cite{molina04a}, such as the case of half filling 
and odd values of $L_\mathrm{S}$. The reliability of the extrapolations of the 
numerically obtained values to infinite lead length becomes limited when the 
energy resolution given by the level spacing in the largest rings is 
insufficient to resolve the resonance at the band center, whose width strongly 
decreases with increasing odd $L_\mathrm{S}$ and $U$. In order to overcome this
problem, considerably larger ring sizes would have to be considered to improve 
the extrapolations. However, it appears impossible within the standard 
embedding method, and with the present computational resources, to perform 
precise enough calculations for sufficiently large ring sizes. In the next 
subsection we develop an improved embedding method that allows to overcome 
these difficulties.

\subsection{Improved embedding method: Damped boundary conditions}

\begin{figure}[t]
\centerline{\includegraphics[width=0.8\columnwidth]{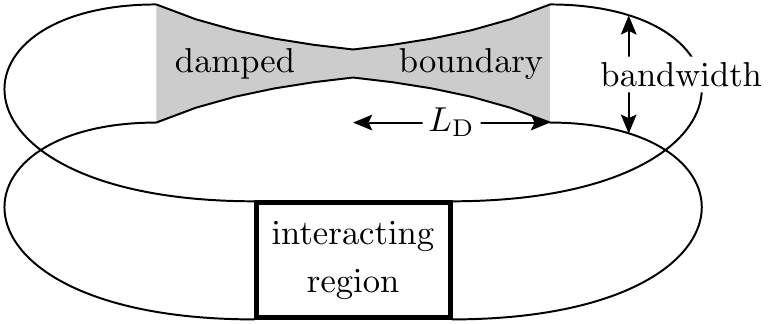}}
\caption{\label{fig:damping}
Sketch of the damped boundary condition setup. The hopping matrix elements are
exponentially reduced on the $L_\mathrm{D}$ last sites on both sides of the 
interacting region, locally suppressing the bandwidth as indicated by the
distance between the upper and lower line. The increased density of
states in the damped boundary region simulates an effectively longer lead.}
\end{figure}

In order to overcome the limitation in ring size, we introduce damped boundary 
conditions \cite{bohr06} that allow to simulate longer effective lead lengths. 
The damped boundary conditions consist in a reduction of the last $L_\mathrm{D}$
hopping elements of the Hamiltonian on both sides of the Hubbard chain, leading
to a region with reduced local bandwidth in the ring opposite to the Hubbard 
part, as indicated in Figure~\ref{fig:damping} by the distance between the
upper and the lower line. The increased local density of 
states is reminiscent of a long lead having full hopping values. We 
choose an exponential reduction controlled by a damping parameter $\Lambda<1$ 
such that the last hopping elements on either side of the interacting region 
are given by $\Lambda^j$ with $j$ counting the last $L_\mathrm{D}$ sites.

\begin{figure}[t]
\centerline{\includegraphics[width=0.9\columnwidth]{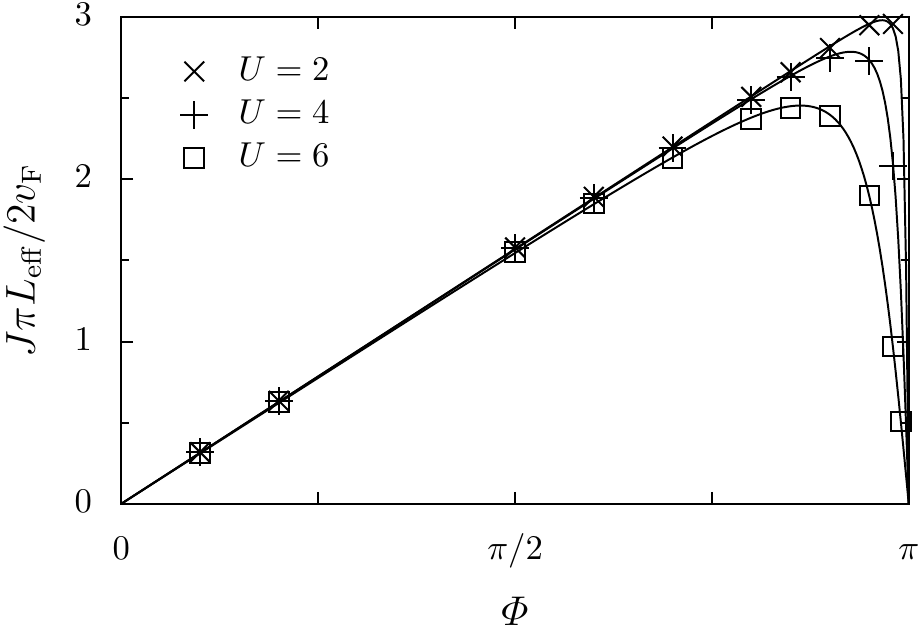}}
\caption{\label{fig:dampedleads}
Persistent currents as a function of flux for half-filled Hubbard rings with 
$L=30$ and $L_\mathrm{S}=3$. Data points are obtained using damped boundary 
conditions with $L_\mathrm{D}=10$ and damping $\Lambda=0.7$. Crosses, pluses, 
and squares are for interaction strengths $U=2$, 4, and 6, respectively. The 
lines are two-parameter fits of the formula \eqref{gogolin-pc} yielding the 
values of Table~\ref{table-Ls3} for the transmission and effective lengths 
$L_\mathrm{eff}=278$, 282, and 295, respectively.}
\end{figure}

We have checked that for the cases of Figure~\ref{fig:g-hubbard} the results
using damped boundary conditions reproduce those from the reliably extrapolated
standard embedding method. However, the usefulness of the improved method
appears in the cases previously discussed where the standard embedding method
meets its numerical limitations. We therefore show in
Figure~\ref{fig:dampedleads} numerical results for the persistent currents
using damped boundary conditions for $L_\mathrm{S}=10$, with $L_\mathrm{D}=10$
and $\Lambda=0.7$. The lines are fits of the analytical result for the
persistent current of noninteracting particles without damped boundary
conditions (formula \eqref{gogolin-pc} with an additional factor of 2 for the
spin). The transmission of the scatterer and the effective length of the ring
have been used as fitting parameters. The quality of the fits is a strong
indication for the validity of the embedding method in the Hubbard model case
since it provides a comparison of the flux dependence with that of the
noninteracting case as it was done for the single-channel case
\cite{molina04a}.

The fitting parameters show indeed that the effective lead length can
increase considerably with the damping, provided it is not chosen too
strong. Damping parameters larger than $\Lambda=0.7$ result, for the system
parameters chosen here, in backscattering, and therefore render the
reliability of the extracted conductances questionable. Remarkably, the
resulting transmission approaches the expected perfect conductance (see
Table~\ref{table-Ls3}). Thereby we have obtained a considerable improvement
from the standard extrapolation procedure, in particular for cases where
the latter yields poor results.  The results of the improved embedding
method can thus be considered to be consistent with the perfect conductance
for all odd values of $L_\mathrm{S}$ that is expected from perturbative
arguments \cite{oguri99,oguri01} and NRG \cite{oguri05}.

\section{Persistent currents for two degenerate channels}
\label{section_pcftcwrcm}

In the previous section we considered the case of degenerate and uncoupled
channels, where the generalization of the one-dimensional embedding method is 
rather straightforward. Once we take into account channel mixing, the 
one-to-one correspondence between persistent currents and transmission
coefficients is lost. However, we still expect that the $\lambda$ parameters of
the polar decomposition \eqref{MS} will be determining the 
corresponding persistent currents. We will treat in this section an arbitrary 
channel mixing, and in order to keep the problem tractable we consider 
quasi-degenerate channels, where $|k_1-k_2| \ll k$. In this case, if we neglect
the weak $k$-dependence of $\tens{u}$, $\tens{v}$, and 
$\tens{\lambda}$, the solutions of \eqref{ring-transfer_3} are 
doubly degenerate, since the r{\^o}les of $A_1^{(p)}$ and  $A_2^{(p)}$ (or 
$B_1^{(p)}$ and  $B_2^{(p)}$) are interchangeable. 

\subsection{Symmetric nanosystems with fixed channel mixing}

In the case where the nanosystem exhibits left-right inversion symmetry we
can set $\alpha=\epsilon$, $\beta=\eta$, $\gamma=\theta$, and
$\varphi=\psi$ in \eqref{u1v1lambda1}. This assumption considerably
simplifies the expression of $\text{Tr}\{\tens{M}\}$ and $F(\tens{M})$
presented in  Appendix~\ref{appendix_ftchp}, and thus the resulting
analytical calculations.

Even in the symmetric case the problem is still rather complicated. We 
therefore consider particular values of $\varphi$. For instance, for $\varphi=0$, 
\eqref{ring-transfer_3} leads to 
\begin{equation}
\cos(2[kL_\mathrm{L}+\alpha+\beta+\gamma])=\cos(2[\alpha-\beta-\gamma])
\end{equation}
and one gets two families of solutions
\begin{equation}
\begin{aligned}
k^{(\ell,1)}L_\mathrm{L}=&-2\alpha+\ell \pi\\
k^{(\ell,2)}L_\mathrm{L}=&-2(\beta+\gamma)+\ell \pi\, .\\
\end{aligned}
\end{equation}
Using \eqref{per_curr_2}, these quantized $k$-values lead to the 
single-levels currents 
\begin{equation}
j^{(\ell,b)}=(-1)^{\ell}\frac{e\hbar}{mL_\mathrm{L}}k^{(\ell,b)}|t_{b}|\quad
b=1,2\,,
\end{equation}
Choosing $\varphi=\pi/2$, one gets similar results with an interchange of $t_1$ 
and $t_2$, and a relative sign between the two current contributions. The fact 
that the persistent currents associated with the ring eigenmodes are solely 
determined by the transmission eigenvalues is to be expected since for these 
particular values of $\varphi$ there is no mode-mixing by 
$\tens{M}$. 

A more interesting case is that of $\varphi=\pi/4$, which corresponds to the 
maximum mode-mixing and the most likely value if we take the matrix
$\tens{u}$ as uniformly distributed in the unitary ensemble 
\cite{been_97}. The corresponding quantization condition is
\begin{equation}
k^{(\ell,b)}L_\mathrm{L}=-\alpha-\beta-\gamma+\frac{(-1)^{b}}{2}
{\rm arccos}(\chi)+ \ell \pi 
\end{equation}
with
\begin{align}
\chi&=\Delta\sin^2(\alpha-\gamma)+\cos^2(\alpha-\gamma) \cos(2\beta) 
\, , \\
\Delta&=t_1t_2-r_1r_2  \, .
\end{align} 
The resulting persistent currents are given by 
\begin{equation}\label{eq:per_curr_pio4}
\begin{aligned}
j^{(\ell,b)}=&(-1)^{\ell+1}\frac{e\hbar k^{(\ell,b)}}{2mL_\mathrm{L}}
\mathrm{sgn}[\cos(\alpha-\gamma)]\\
\times&
\left\{\frac{(t_2-t_1)\mathrm{sgn}[\cos(\beta)]}
            {\sqrt{1+\frac{1+\Delta}{2\cos^2(\beta)}\tan^2(\alpha-\gamma)}}
             \right.\\
&\left. +(-1)^b\frac{(t_1+t_2)\mathrm{sgn}[\sin(\beta)]}
            {\sqrt{1+\frac{1-\Delta}{2\sin^2(\beta)}\tan^2(\alpha-\gamma)}}
\right\} \, ,
\end{aligned}
\end{equation}
and therefore are bounded by
\begin{equation}
\left|j^{(\ell,b)}\right|
\le\frac{e\hbar k^{(\ell,b)}}{mL_\mathrm{L}}\max\{t_1,t_2\}\, .
\end{equation}
In the limit of large $L_\mathrm{L}$ we can take 
$k^{(\ell,1)}\simeq k^{(\ell,2)}\simeq k^{(\ell)}=\ell\pi/L_\mathrm{L}$ 
and show that
\begin{align}\label{eq:condition_a} 
\left|j^{(\ell,1)}+j^{(\ell,2)}\right| &\le 
\frac{e\hbar k^{(\ell)}}{mL_\mathrm{L}}\left|t_1-t_2\right|\, ,  \\
\left|j^{(\ell,1)}-j^{(\ell,2)}\right| &\le 
\frac{e\hbar k^{(\ell)}}{mL_\mathrm{L}}\left(t_1+t_2\right)\, .
\label{eq:condition_b}
\end{align}

\begin{figure}
\centerline{\includegraphics[width=0.9\columnwidth]{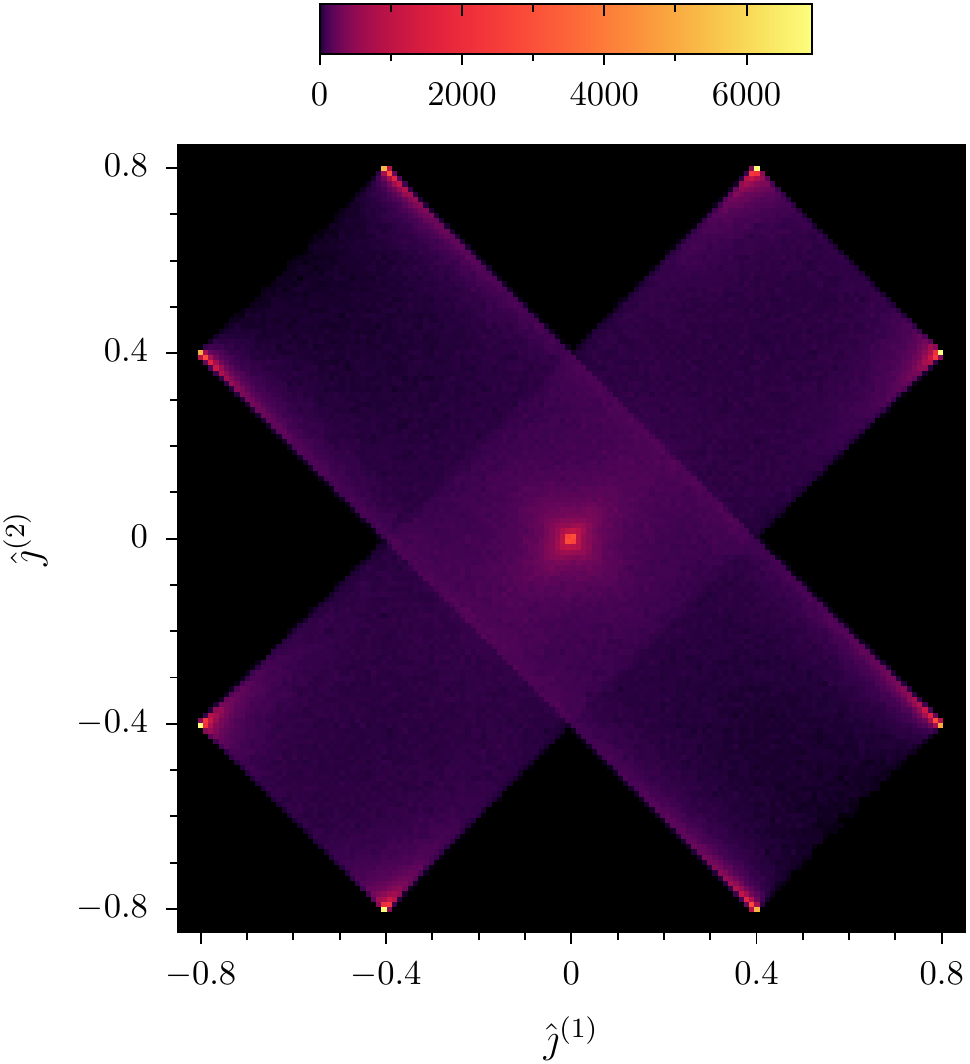}}
\caption{Color scale plot of the distribution 
$(\hat{\jmath}^{(1)},\hat{\jmath}^{(2)})$ for $t_1=0.8$ and $t_2=0.4$ in the
case of a left-right symmetric scatterer with a channel mixing angle 
$\varphi=\pi/4$. 
\label{fig:j1j2piover4}}
\end{figure}

We remark that, leaving aside the trivial case of uncoupled channels, the 
relationship between persistent current and transmission amplitudes is not 
unique, as it involves the mixing angles. However, already the special case
$\varphi=\pi/4$ shows that there are stringent constraints on the possible
values of $t_1$ and $t_2$. If we disposed of an ensemble of different 
scatterers with fixed $t_1$, $t_2$, and mixing angle $\varphi$, we could 
extract the transmission amplitudes by  varying the angles $\alpha$, $\beta$ 
and $\gamma$ through the ensemble. We carry on this procedure by choosing these
three angles uniformly distributed in the interval $(0,2\pi)$, solving the 
resulting eigenvalue problem \eqref{ring-quant_cond}, and then obtaining from 
\eqref{persistent_current} the two resulting persistent currents. For 
$\varphi=\pi/4$ we obtain the distribution of normalized persistent currents
$\hat{\jmath}^{(1,2)}=mL_\mathrm{L}/(e\hbar |k|)j^{(1,2)}$ shown in 
Figure~\ref{fig:j1j2piover4}, where we have arbitrarily taken $t_1=0.8$ and 
$t_2=0.4$. The restrictions imposed by \eqref{eq:condition_a} and 
\eqref{eq:condition_b} constrain the possible values of
$(\hat{\jmath}^{(1)},\hat{\jmath}^{(2)})$ to the rectangle defined by
$(\pm t_{1,2},\mp t_{2,1})$. The rectangle $(\pm t_{1,2},\pm t_{2,1})$ appears 
because the chosen pairs $(\hat{\jmath}^{(1)},\hat{\jmath}^{(2)})$ do not
necessarily have the same $\ell$, as was the case in \eqref{eq:per_curr_pio4}.
Interestingly, the $(\hat{\jmath}^{(1)},\hat{\jmath}^{(2)})$ distribution is 
quite non-uniform, but concentrated around $(0,0)$ and, in particular, at the 
vertices of the rectangles.

\begin{figure}
\centerline{\includegraphics[width=0.9\columnwidth]{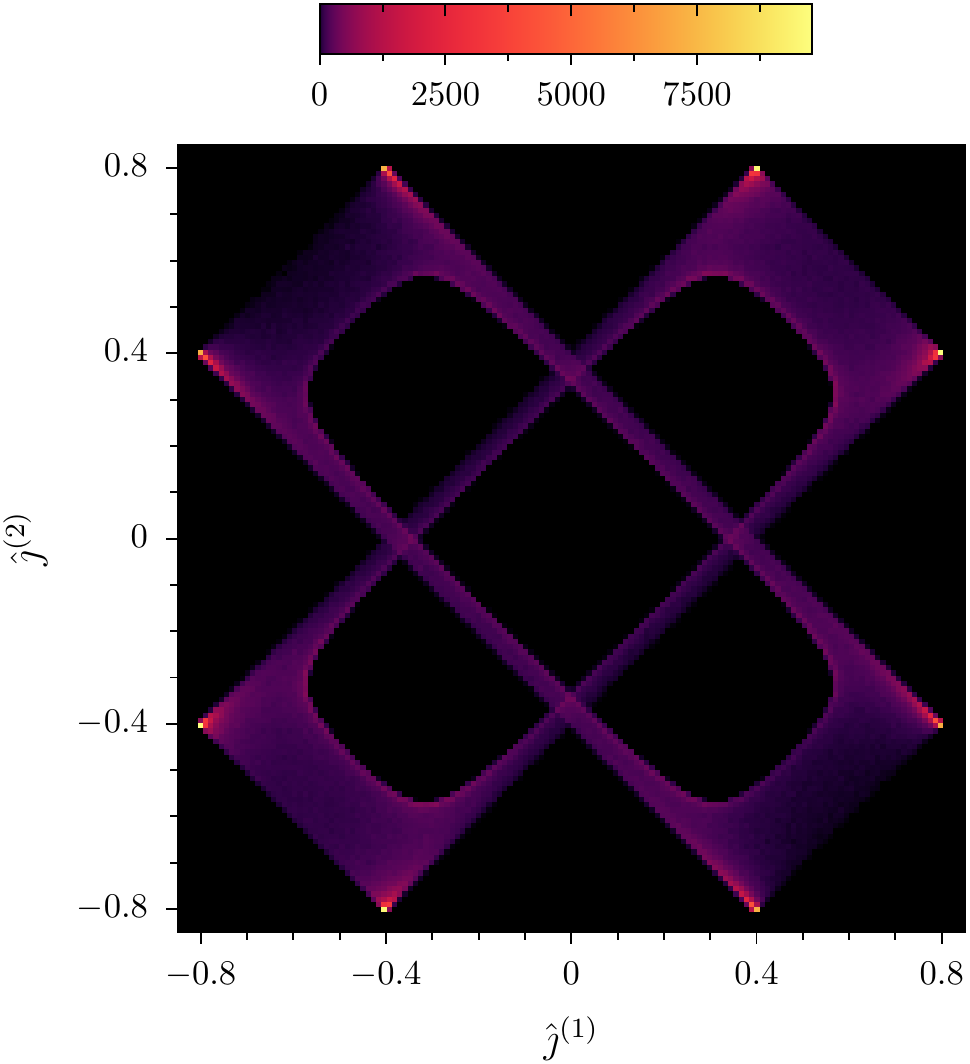}}
\caption{Color scale plot of the distribution
$(\hat{\jmath}^{(1)},\hat{\jmath}^{(2)})$ for $t_1=0.8$ and $t_2=0.4$ in the
case of a left-right symmetric scatterer with a channel mixing angle 
$\varphi=\pi/8$.
\label{fig:pi8}}
\end{figure}

Other mixing angles are difficult to deal with at the analytical level, but the
$(\hat{\jmath}^{(1)},\hat{\jmath}^{(2)})$ distribution can be obtained as in the
previous case. This is done in Figure~\ref{fig:pi8} for $\varphi=\pi/8$, and with
the same values of $t_1$ and $t_2$ as before. We see that the restrictions
\eqref{eq:condition_a} and \eqref{eq:condition_b} still apply, the distribution
is depleted around the central point $(0,0)$, and the concentration around the
vertices of the rectangles is more pronounced than for the case of 
$\varphi=\pi/4$. In Figure~\ref{fig:j1j2versusphi} we present the probability 
distribution of $(\hat{\jmath}^{(1)},\hat{\jmath}^{(2)})$ as a function of the 
angle $\varphi$. Considering different horizontal cross-sections, like 
Figures~\ref{fig:j1j2piover4} and \ref{fig:pi8}, we can follow the evolution from
the uncoupled channel case to that of intermediate mixings.

\subsection{Asymmetric nanosystems and random channel mixing}

In the previous subsection we have seen that in the case of left-right 
inversion symmetry the transmission amplitudes can be inferred from the study 
of the distribution of matrices $\tens{M}$ with fixed radial
parameters $\lambda_{1,2}$ and varying angular parameters. This procedure can 
be formalized by taking a random matrix approach for the $\tens{u}$ 
and $\tens{v}$ matrices, while keeping $\tens{\Gamma}$ 
fixed. This kind of approach is widely used in the application of random matrix
theory to quantum transport, where we separate the statistics of the matrices 
$\tens{u}$ and $\tens{v}$  from that of the transmission eigenvalues. While
the first ones are assumed to be uniformly distributed over the unitary
ensemble, the distribution of the latter depends on the nature of 
the problem under study, i.e.\ its diffusive or chaotic character \cite{jal_95}. 
In practice, such a method would require to connect 
the nanosystem under study to mode-mixing reflectionless diffusors, represented 
by matrices uniformly distributed on the unitary ensemble, and to study the 
resulting distribution of persistent currents.

\begin{figure}
\centerline{\includegraphics[width=0.9\columnwidth]{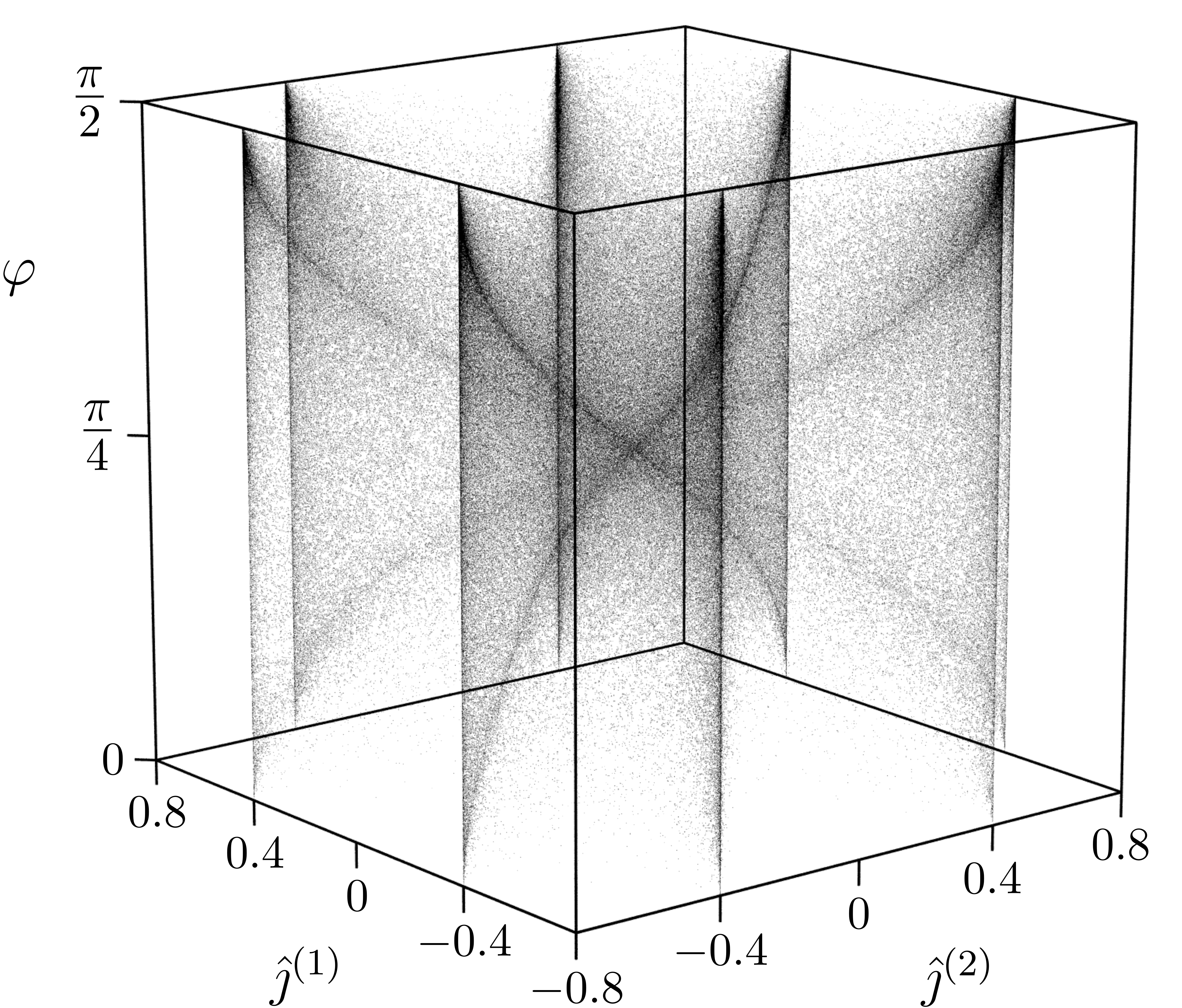}}
\caption{Probability density to find a pair 
$(\hat{\jmath}^{(1)},\hat{\jmath}^{(2)})$ for $t_1=0.8$ and $t_2=0.4$ with 
left-right symmetry as a function of the mixing angle $\varphi$ 
(vertical axis). 
\label{fig:j1j2versusphi}}
\end{figure}
\begin{figure}
\centerline{\includegraphics[width=0.9\columnwidth]{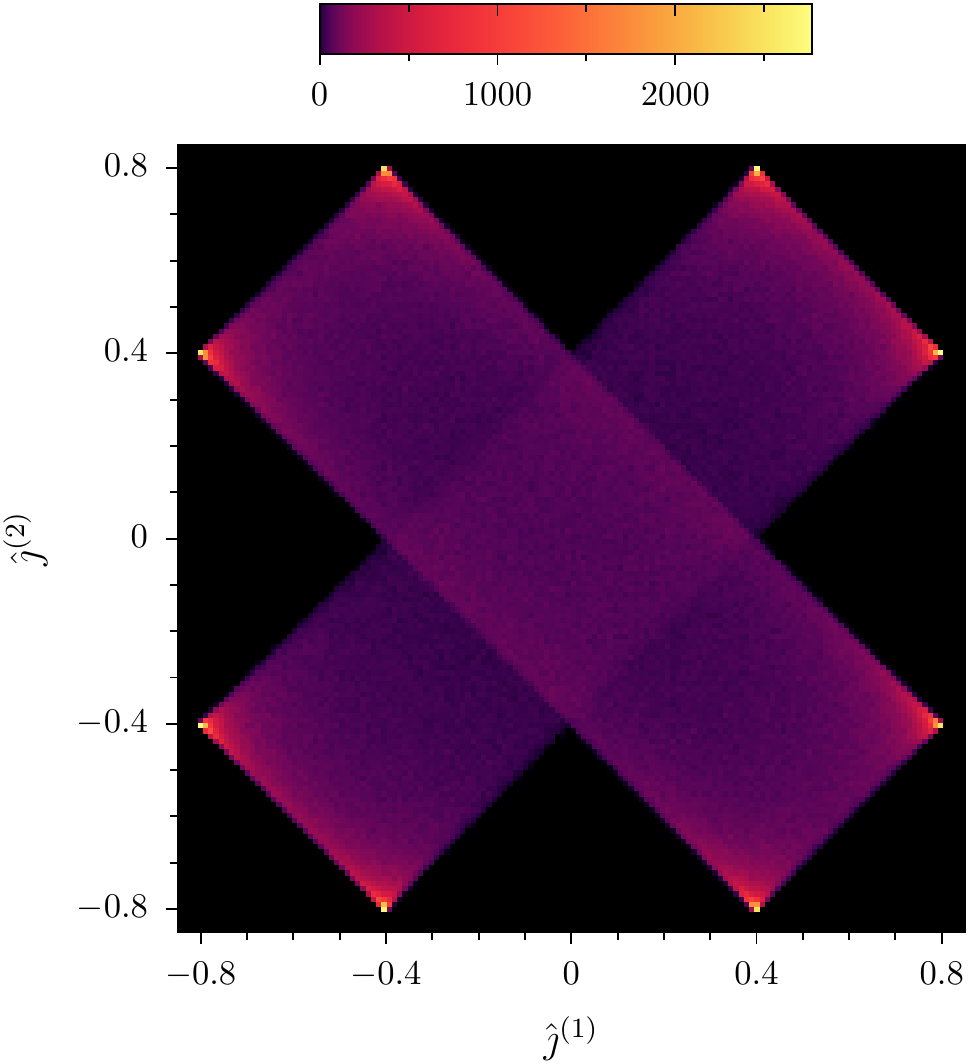}}
\caption{Color scale plot of the probability density to find a pair 
$(\hat{\jmath}^{(1)},\hat{\jmath}^{(2)})$ for $t_1=0.8$ and $t_2=0.4$ with 
random channel mixing. 
\label{fig:distrib-j1_j2}}
\end{figure}

Since our procedure will be carried out numerically, we can drop the symmetry 
requirement of the previous subsection. We then generate a large number of 
independent unitary matrices $\tens{u}$ and $\tens{v}$ 
for fixed transmission amplitudes $t_1$ and $t_2$. The two non-degenerate 
$k$-values arising from the quantization condition \eqref{ring-transfer_3} for 
$\Phi=\pi/2$ yield pairs of normalized persistent currents $\hat{\jmath}^{(1)}$
and $\hat{\jmath}^{(2)}$, that we obtain from Equation~\eqref{persistent_current}. 
The resulting distribution is presented in Figure~\ref{fig:distrib-j1_j2}. 
Comparing with Figure~\ref{fig:j1j2versusphi}, there are some differences since 
we do not have spatial symmetry and the angle $\varphi$ is averaged over 
the unitary ensemble. However, similar information can be extracted from both 
cases: the transmission amplitudes appear as the bounds and the most likely 
values of the $(j^{(1)},j^{(2)})$ distribution.

\begin{figure}
\centerline{\includegraphics[width=0.9\columnwidth]{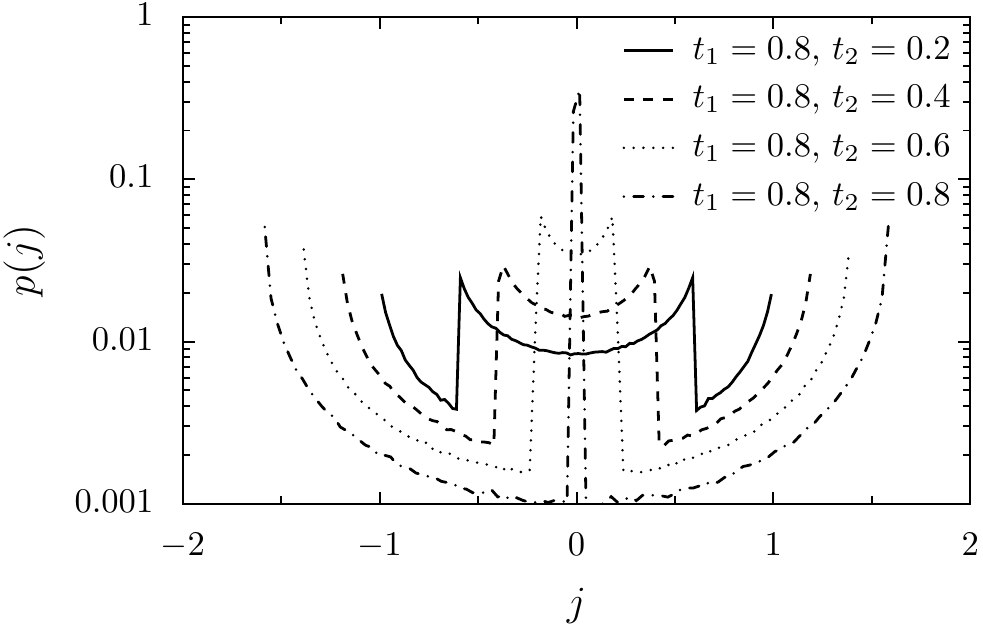}}
\caption{\label{fig:distrib-2}
Probability distribution $p(\hat{\jmath})$ of the normalized total persistent 
current $\hat{\jmath}$ in a ring containing a two-channel scatterer with 
random channel mixing for $t_1=0.8$ and $t_2=0.2$ (solid line), $0.4$ (dashed),
$0.6$ (dotted) and $0.8$ (dashed-dotted).}
\end{figure}

If we now consider a non-interacting many-particle system, the sum over the 
occupied levels should be carried out in order to obtain the total persistent 
current. Assuming that, like in the $N=1$ case, the total persistent current is
dominated by the contribution at the Fermi level, it is of interest to consider
the distribution of the sum 
$\hat{\jmath}=\hat{\jmath}^{(1)}+\hat{\jmath}^{(2)}$. Such a distribution is 
presented in  Figure~\ref{fig:distrib-2} on a logarithmic scale for $t_1=0.8$ and 
various values of $t_2\le t_1$. The structure found in 
Figure~\ref{fig:distrib-j1_j2} translates into bounds of $\hat{\jmath}$ at 
$\pm (t_1+t_2)$ and strong peaks at  $\pm |t_1-t_2|$. Therefore, from the 
distribution of $\hat{\jmath}$ when branching the nanosystem to random 
reflectionless diffusors we can infer the values of the amplitudes $t_1$ and 
$t_2$, and eventually the conductance of the system.

The relationships that we have found between persistent currents and
transmission amplitudes in the non-interacting two-channel case are 
interesting, but difficult to use in a practical way. The extension beyond 
$N=2$ is in principle possible. However, as $N$ grows it becomes increasingly 
difficult to identify the $N$ transmission eigenvalues from the distribution of
$\hat{\jmath}=\sum_{b=1}^{N}\hat{\jmath}^{(b)}$. Our main interest is in the 
$N=2$ case once the interactions are switched on. The use of the statistical 
approach for the interacting case, leaving practical considerations aside, 
would rely on three key  assumptions. Firstly, like in the $N=1$ case, we have 
to accept that the interacting problem is described by effective one-particle 
scattering parameters. Secondly, the total persistent current should be 
dominated by its Fermi level contribution. Lastly, the two channels at the 
Fermi level need to be quasi-degenerate. 

\section{Reduction of multi-channel scattering to single-channel problems}
\label{sec:reduction}

As we have seen in the previous section, it is not obvious how to extract the 
transmission probabilities of an $N$-channel scatterer required for the 
calculation of the conductance from the persistent current distribution of 
an $N$-channel ring. However, it is possible to calculate several 
single-channel persistent currents which then allow in principle to 
characterize completely the many-channel scattering matrix.

\begin{figure}
\centerline{\includegraphics[width=\columnwidth]{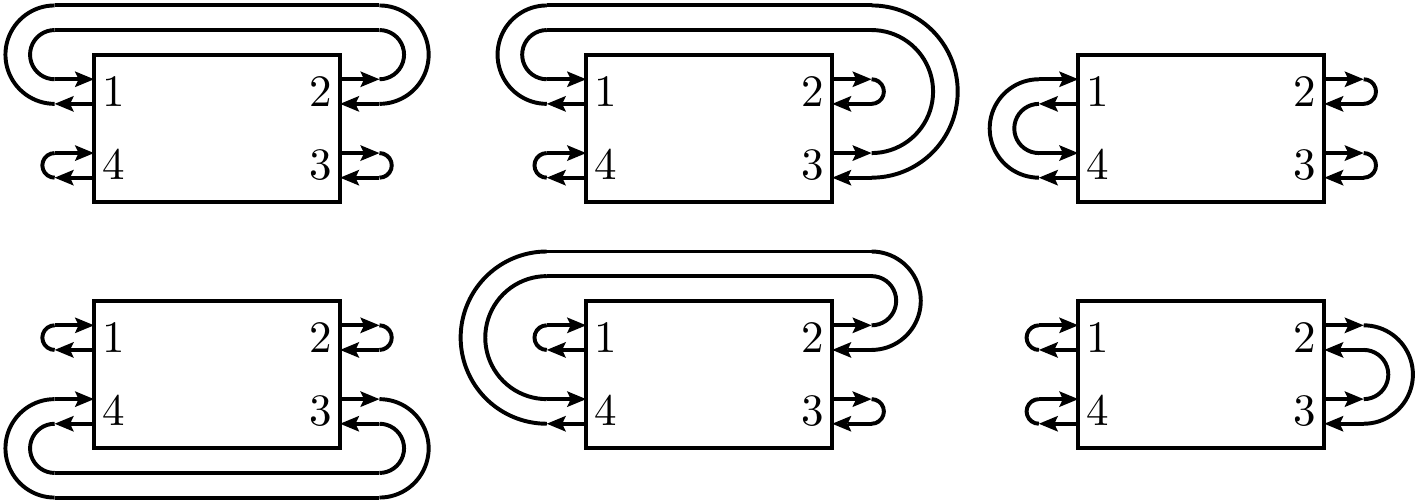}}
\caption{\label{fig:one-ring}For a two-channel scattering problem there exist
six different one-channel scattering problems which can be treated with the
standard embedding method.}
\end{figure}

The effective one-channel scattering problem $ij$ is obtained by connecting
two channels $i$ and $j$ which both can be situated on either side of the
scattering region. All other channels are closed by reinjecting the
outgoing current into the scattering region. For the example of two
channels we present in Figure~\ref{fig:one-ring} the six different possible
setups. Note that the noninteracting ring does not necessarily connect
opposite sides of the interacting region. In principle, the electrons can
acquire an arbitrary phase upon injection. While this phase may represent a
useful tool, we will set it to zero in the sequel. For an $N$-channel
problem one can construct in this way $2N^2-N$ different one-channel
scattering problems. 

Within the embedding method, the effective one-channel setup linking channels
$i$ and $j$ has an associated persistent current from which the modulus 
$T_{ij}$ of the corresponding transmission amplitude can be determined. 
Clearly, the $T_{ij}$ are functions of the $2N^2+N$ parameters defining the
matrices $\tens{u}$, $\tens{v}$, and
$\tens{\lambda}$. Such functions are presented in the appendix for
the two-channel case. In the general case, we have $2N^2-N$ different $T_{ij}$,
so that $2N$ irrelevant phases are left undetermined.

Our goal is to extract the radial parameters $\lambda_a$ that determine the
conductance. Already the application of the embedding method to all of the
$2N^2-N$ one-channel problems may require considerable numerical resources and
the procedure might be quite lengthy. In addition, the complicated functional
dependence of the $T_{ij}$ makes obtaining the $\lambda_a$ a challenging task.
Depending on the symmetries of the system under investigation, both the number
of parameters appearing in the transfer matrix and the number of nonequivalent
one-channel problems may be reduced with respect to these general expressions.

Furthermore, interaction-induced nonlocal effects impose some restrictions. For
the effective transmission of interacting one-dimensional scatterers, it has 
been found that the usual composition law of scatterers in series does not hold
\cite{molina05,weinmann08}. The deviations can be interpreted as a nonlocal 
interaction-induced effect on the effective transmission of an interacting 
scatterer. This effect disappears only in the limit of large distance between 
the scatterers. In the present situation, the closing of channels might 
represent a perturbation leading to such a nonlocal effect unless the 
connecting leads are sufficiently long. 

Despite these difficulties the proposed method might provide a possible 
approach for the demanding problem of many-body transport with mixing channels.
For instance, we might study larger Hubbard chains than the ones we have 
addressed with the other methods. In addition, the fact of separating the 
problem into single-channel setups can be useful for the general analysis of 
the multi-channel scattering case as will be exemplified in 
Appendix~\ref{sec:scsappl}.

\section{Conclusions}

The embedding method for extracting the conductance of an interacting
nanosystem is by now a well established and widely used procedure in the case
of one-dimensional spinless fermions
\cite{favand98,sushkov01,molina03,meden03,rejec03,molina04a,molina05}. Its
validity has been ascertained and interesting physical phenomena have been
found from its application \cite{molina04a,molina05,vasseur_thesis}. The
extension of the embedding method to the case of two channels is therefore
of great importance.  In this work we have presented various possible
extensions of the $N=1$ embedding method to $N=2$, depending on the nature
of the channels.

When the channels represent the two spin directions, we have a
one-dimensional Hubbard chain, where the absence of an effective spin-flip
term allows for a straightforward generalization of the $N=1$ embedding
method. The use of the DMRG algorithm in the presence of spin makes the
problem numerically demanding, but we have checked that the method is
reliable for chains of up to 4 interacting sites. We have introduced an
improvement of the standard embedding method by considering damped boundary
conditions, allowing to address larger systems and to deal with resonances, 
which is the case for an odd number of interacting sites $L_\mathrm{S}$. Our 
results are compatible with the perfect conductance predicted for a Hubbard 
chain with an odd number of sites \cite{oguri99}. For even $L_\mathrm{S}$ the
conductance is reduced by the electron-electron interactions. Similar
qualitative behavior has been established for Hubbard chains coupled to
reservoirs using NRG \cite{oguri05}.

The generalization of the embedding method to the case of two mixed
channels is difficult due to the increased number of transmission
amplitudes that are necessary to determine the conductance. In the case of
quasi-degenerate channels, by connecting different pairs of channels we
reduce the problem to various one-dimensional setups, where the
corresponding transmission amplitude can be obtained from the associated
persistent current. The implementation of such a method requires
considerable numerical resources and solving a system of eight coupled
non-linear equations.  In the non-interacting case with quasi-degenerate
channels we proposed a statistical approach that leads to the values of the
transmission amplitudes from the distribution of persistent currents
obtained while varying the mixing of the channels. The practical
implementation of this method in order to extract the conductance of a
given nanosystem would rely in coupling it with an ensemble of
reflectionless diffusors that scramble the phases of the transfer matrix
without affecting the transmission amplitudes.

The embedding method is based on an
effective one-particle description of a correlated system. Anderson-like
quantum impurity models like that of a short Hubbard chain scattering
conduction electrons or the Interacting Resonant Level Model
\cite{mehta06,doyon07,boulat08} have been studied in the
literature through the numerical renormalization group (NRG) 
\cite{hewson_book,oguri05}
or the Bethe-Ansatz \cite{mehta06,doyon07,boulat08}. From these studies we know 
that the low-energy excitations behave like free fer\-mi\-ons in the 
zero-temperature limit of those models. This is the case in r\'egimes where the 
effect of interactions remains perturbative, as well as in the non-perturbative
limit. In this latter limit, where (spin or orbital) Kondo effects occur, the 
extended conduction electrons continue to behave as free fermions below the 
Kondo temperature, as first shown by Wilson \cite{hewson_book}.

Extending this concept to arbitrary interacting nanosystems, one can justify 
the assumption that an interacting region embedded between non-interacting 
leads can still be described at sufficiently low temperatures and sufficiently 
far away as an effective one-body scatterer with interaction-dependent 
effective parameters. In certain studies of quantum impurity models, the 
effective quantum transmission is extracted directly from the knowledge of the 
effective one-body excitations or, via the Friedel sum rule, from the knowledge
\cite{ng88,freyn10} of the occupation numbers of the interacting region.  

Although, not surprisingly, many-channel problems require a larger 
numerical effort, the extension of the embedding method to more than one
channel is feasible and we have presented several ideas to achieve this
generalization. The embedding method is thus available for future practical 
applications beyond the one-channel case.

\begin{acknowledgement}
We thank R.A.\ Molina for useful discussions and a careful reading of the
manuscript and H.U.\ Baranger for clarifications about the symmetries in the
Hubbard problem. R.A.J.\ thanks the Institute for Nuclear Theory at the
University of Washington for its hospitality and the Department of Energy for
partial support during the completion of this work. Financial support from the
European Union within the MCRTN program and from the ANR (grant
ANR-08-BLAN-0030-02) are gratefully acknowledged.
\end{acknowledgement}

\appendix

\section{Formulas for the two-channel problem}
\label{appendix_ftchp}

In this appendix we present the explicit form of 
$\text{Tr}\{\tens{M}\}$ and $F(\tens{M})$, which are 
necessary for solving  the quantization condition \eqref{ring-transfer_2} of 
the two-channel problem. We are interested in expressing them in terms of 
$k_1 \pm k_2$. For the former we have
\begin{equation}\label{Tr_M}
\begin{aligned}
\text{Tr}\{\tens{M}\} = T^{1/2}\big[
&\cos(k_1 L_\mathrm{L}-\xi_1)\cos(\Omega)\\
&+\cos(k_2 L_\mathrm{L}-\xi_2)\sin(\Omega)\big]
\, ,
\end{aligned}
\end{equation}
with
\begin{equation}\label{T}
\begin{aligned}
\frac{T}{4} = &\frac{1+\cos(2\varphi)\cos(2\psi)}{2}
\left(\frac{1}{t_1^2}+\frac{1}{t_2^2}\right) \\
&-\frac{1}{t_1t_2}\sin(2\varphi)\sin(2\psi)\cos(\beta+\eta)
\, ,
\end{aligned} 
\end{equation}
\begin{equation}\label{xi_1}
\begin{aligned}
\tan{\xi_1}&=\\
&\hspace{-3.2pt}\frac{t_1\tan(\varphi)\tan(\psi)\cos(\alpha+\beta+\epsilon+\eta)
-t_2\cos(\alpha+\epsilon)}
{t_1\tan(\varphi)\tan(\psi)\sin(\alpha+\beta+\epsilon+\eta)
-t_2\sin(\alpha+\epsilon)}
\, ,
\end{aligned} 
\end{equation}
\begin{equation}\label{xi_2}
\begin{aligned}
\tan{\xi_2}&=\\
&\hspace{-3.6pt}\frac{t_1\cos(\gamma+\beta+\theta+\eta)
-t_2\tan(\varphi)\tan(\psi)\cos(\gamma+\theta)}
{t_1\sin(\gamma+\beta+\theta+\eta)
-t_2\tan(\varphi)\tan(\psi)\sin(\gamma+\theta)}
\, ,
\end{aligned} 
\end{equation}
\begin{equation}\label{omega}
\begin{aligned}
\tan^2{\Omega}&=\\
&\hspace{-28.5pt}\frac{t_1^2+t_2^2\tan^2(\varphi)\tan^2(\psi)-
2t_1t_2\tan(\varphi)\tan(\psi)\cos(\beta+\eta)}
{t_1^2\tan^2(\varphi)\tan^2(\psi)+t_2^2-
2t_1t_2\tan(\varphi)\tan(\psi)\cos(\beta+\eta)}
\, .
\end{aligned} 
\end{equation}

The evaluation of \eqref{F_subdeterminants} yields 
\begin{equation}\label{eq_F}
\begin{aligned}
\hspace{21pt}&\hspace{-21pt}
F(\tens{M})=\frac{1+\cos(2\varphi)\cos(2\psi)}{2}\\
&- \frac{\sin(2\varphi)\sin(2\psi)[\cos(\beta+\eta)-
r_1r_2\cos(\beta-\eta)]}{2t_1t_2}\\
&+\frac{\cos([k_1+k_2]L_\mathrm{L}+\alpha+\beta+\gamma+\epsilon+\eta+\theta)}
{t_1t_2}\\
&+\frac{\sin(2\varphi)\sin(2\psi)
\cos([k_1-k_2]L_\mathrm{L}+\alpha+\epsilon-\gamma-\theta)}{2}\\
&-\frac{1}{2t_1t_2}\big(\cos([k_1-k_2]L_\mathrm{L}+
\alpha+\epsilon-\gamma-\theta)\\
&\quad \times [(1+\cos(2\varphi)\cos(2\psi))
\cos(\beta+\eta)\\
&\qquad +r_1r_2(1-\cos(2\varphi)\cos(2\psi))\cos(\beta-\eta)]\\
&+\sin([k_1-k_2]L_\mathrm{L}+\alpha+\epsilon-\gamma-\theta)\\
&\quad \times[(\cos(2\varphi)+\cos(2\psi))\sin(\beta+\eta)\\
&\qquad +r_1r_2(\cos(2\varphi)-\cos(2\psi))\sin(\beta-\eta)] 
\big)\,.
\end{aligned} 
\end{equation}

The above expressions simplify considerably in the case of left-right inversion
symmetry, allowing to undertake the analytic calculations of 
Section~\ref{section_pcftcwrcm}. However, this approximation leaves aside 
important cases, like the one of disordered systems.

\section{Alternative proof of the spin conservation using the 
reduction of multichannel scattering to single-channel problems}
\label{sec:scsappl}

We now demonstrate that the results \eqref{tupdown0} and \eqref{tdiag} can
alternatively be obtained by means of the single-channel scattering problems
introduced in Section~\ref{sec:reduction}. To this end, we show that in four of 
the six scattering configurations the persistent current vanishes.
   
The number of electrons $N_\sigma=\sum_i \hat{n}_{i,\sigma}$ having spin 
$\sigma$ commutes with the Hubbard Hamiltonian \eqref{eq:hamiltonian}.
Therefore, in a closed system the number of electrons having spin up or spin 
down cannot change.  While the total spin of the electrons inside the 
interacting nanosystem can fluctuate when electrons move between the leads and 
the nanosystem, in a situation of stationary transport through the system, the 
mean spin of the electrons entering the system has to equal the mean spin of 
the electrons leaving the nanosystem. It is then possible to describe the 
effective stationary transmission through the Hubbard chain without invoking 
transmission processes that are accompanied by a spin flip. 

\begin{figure}
\centerline{\includegraphics[width=0.6\columnwidth]{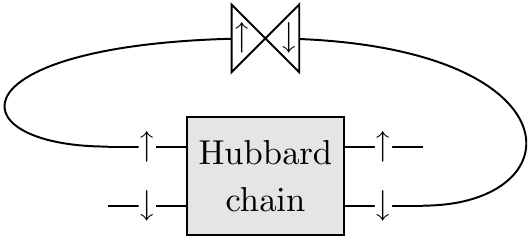}}
\caption{\label{sketch-hubbard-ring}A Hubbard chain is closed to a ring by a 
single-channel lead attached to different spin channels. Within the embedding 
method, one therefore has to introduce a spin-flip in the noninteracting lead.}
\end{figure}

The absence of spin-flip transmissions \eqref{tupdown0} can be confirmed using 
the formalism presented in Section~\ref{sec:reduction}. Four of the six 
possible single-channel configurations connect channels of different spin 
orientation. The spin has then to be flipped somewhere in the noninteracting 
lead as is indicated in Figure~\ref{sketch-hubbard-ring} for one of these 
configurations. The contribution of the kinetic energy to the total Hamiltonian
\eqref{eq:hamiltonian} in this example reads
\begin{equation}\label{hubbard-onechannel}
\begin{aligned}
H_\mathrm{K} =- \Bigg[
\sum_{i=-L/2}^{L_\mathrm{S}-1} 
c_{i,\uparrow}^\dagger c_{i+1,\uparrow}^{\phantom{\dagger}} 
&+\sum_{i=1}^{L/2-1} 
c_{i,\downarrow}^\dagger c_{i+1,\downarrow}^{\phantom{\dagger}}\\
&+c_{-L/2,\uparrow}^\dagger c_{L/2,\downarrow}^{\phantom{\dagger}} 
+ \text{h.c.}\Bigg]\,,
\end{aligned} 
\end{equation}
where the third term on the right-hand side is responsible for the spin flip.

In a single-channel configuration linking different spin channels, no
persistent current can occur. A non-vanishing persistent current in the 
presence of a spin flip term would result in an accumulation of one spin 
orientation in the ring. In a stationary situation, this is of course 
impossible.

It can be seen from the relations presented in Appendix~\ref{appendix_eocp}
between the persistent currents in different single-channel rings and the 
effective transmission amplitudes that the vanishing of persistent currents in
setups connecting different spin channels is consistent with the vanishing of
the transmission amplitudes with spin flip \eqref{tupdown0}.

\section{Reduction of a two-channel scattering problem to effective
one-channel scattering problems} 
\label{appendix_eocp}

As illustrated in Figure~\ref{fig:one-ring} for two channels, the effective
one-channel scattering problem is obtained by closing on themselves one pair of
channels and connecting the remaining pair by a non-interacting lead. The 
closing of a channel allows in \eqref{eq:cd_ms_ab} to make the corresponding 
in- and outgoing amplitudes equal, thus reducing by two the number of 
amplitudes involved in the scattering problem.

The closing of a channel implies that the amplitudes of the corresponding in-
and outgoing channels become equal. Therefore, the closing of two channels 
reduces the number of amplitudes appearing in \eqref{eq:cd_ms_ab} from 8 to 6.
Furthermore, we can use two of these equations in order to eliminate the 
remaining two amplitudes related with the closed channels. This leaves us with
two equations relating the amplitudes of the open channels and thus defining
the effective one-channel problem.

This procedure results in a transfer matrix of considerable complexity when
expressed in terms of the parameters of the full scattering problem. The 
complexity also becomes apparent when one expresses the transfer matrix in 
terms of the transmission and reflection coefficients as well as the phases 
encountered on all internal scattering sequences leading from one open 
channel to the other.

Two different kinds of one-channel setups must be distinguished, depending on 
whether they connect channels on opposite sides or on the same side. For the 
first group the moduli of the transmission amplitudes are
\begin{equation}
\begin{aligned}
  T_{12} &= Q_1(\gamma,\theta,\beta,\eta,\varphi,\psi)\\ 
  T_{34} &= Q_1(\alpha,\epsilon,\beta,\eta,\varphi-\pi/2,\psi-\pi/2)\\
  T_{13} &= Q_1(\gamma,\epsilon,\beta,\eta,\varphi,\psi-\pi/2)\\
  T_{24} &= Q_1(\alpha,\theta,\beta,\eta,\varphi-\pi/2,\psi)\,,
\end{aligned}
\end{equation}
where
\begin{multline}
\label{eq:q1}
Q_1(\omega_1,\omega_2,\omega_3,\omega_4,\omega_5,\omega_6) =\\
    2 \sqrt{\frac{N_1(\omega_1,\omega_2,\omega_3,\omega_4,\omega_5,\omega_6)^2}
    {D_1(\omega_1,\omega_2,\omega_3,\omega_4,\omega_5,\omega_6)}}\,,
\end{multline}
while for the second group we have 
\begin{equation}
\begin{aligned}
  T_{14} &= Q_2(\varphi, \psi, \epsilon,\theta, \eta, \beta)\\
  T_{23} &= Q_2(\psi, \varphi, \alpha, \gamma, \beta, \eta)
\end{aligned}
\end{equation}
with
\begin{multline}
Q_2(\omega_1,\omega_2,\omega_3,\omega_4,\omega_5,\omega_6) =\\
    2 \sqrt{\frac{N_2(\omega_1,\omega_2,\omega_3,\omega_4,\omega_5,\omega_6)^2}
    {D_2(\omega_2,\omega_3,\omega_5,\omega_4)}}\,.
\end{multline}
The fact that each of the two groups is characterized by a single function,
$Q_1$ or $Q_2$, stems from the symmetries of the problem. 

In \eqref{eq:q1}, we have introduced the definitions
\begin{equation}
\begin{aligned}
N_1(\omega_1, \omega_2, \omega_3, \omega_4,\omega_5,\omega_6) &=\\
 t_1 \cos(\omega_5) \cos(\omega_6) 
  [ &r_2 \cos(\omega_1+\omega_3-\omega_2-\omega_4)\\
    &- \cos(\omega_1 + \omega_3 + \omega_2 + \omega_4)]\\
- t_2 \sin(\omega_5) \sin(\omega_6) [&r_1 \cos(\omega_1-\omega_2)\\
                         &- \cos(\omega_1+\omega_2)] 
\end{aligned}
\end{equation}
and
\begin{multline}
    D_1(\omega_1,\omega_2,\omega_3,\omega_4,\omega_5,\omega_6) =\\
    [H_1^\text{c}(\omega_1,\omega_2,\omega_3,\omega_4,\omega_5,\omega_6) - 1]^2\\
    +H_1^\text{s}(\omega_1,\omega_2,\omega_3,\omega_4,\omega_5,\omega_6)^2
\end{multline}
where
\begin{equation}
\begin{aligned}
H_1^{i}(\omega_1,\omega_2,&\omega_3, \omega_4,\omega_5,\omega_6)=\\
r_1& A_2^{i} (\omega_1,\omega_2,\omega_5 -\pi/2,\omega_6 -\pi/2) \\
+ r_2& A_2^{i}(\omega_1+\omega_3, \omega_2 + \omega_4,\omega_5,\omega_6)\\
- r_1r_2& A_3^{i}(\omega_1 + \omega_2 + \omega_3, \omega_1 +
\omega_2 + \omega_4,\omega_5,\omega_6 -\pi/2) \\
+& A_1^{i}
 (\omega_1 + \omega_2, \omega_3 + \omega_4, \omega_5 -\pi/2, \omega_6 -\pi/2)\\
-& A_3^{i}(\omega_1 +\omega_2 + \omega_3 + \omega_4, 
                  \omega_1 + \omega_2, \omega_5, \omega_6)\,,
\end{aligned}
\end{equation}
with $i=\text{c}, \text{s}$ and
\begin{equation}
\begin{aligned}
A_1^\text{c}(\omega_1, \omega_2, \omega_3, \omega_4) =& \\
  2 t_1 t_2 \cos(\omega_3)&\cos(\omega_4)\sin(\omega_3)\sin(\omega_4) 
                                  \cos(2\omega_1+\omega_2) \\
A_1^\text{s}(\omega_1, \omega_2, \omega_3, \omega_4) =& \\
  2 t_1 t_2 \cos(\omega_3)&\cos(\omega_4)\sin(\omega_3)\sin(\omega_4) 
                                  \sin(2\omega_1+\omega_2) \\
A_2^\text{c}(\omega_1, \omega_2, \omega_3, \omega_4) =
 & \cos(\omega_3)^2\cos(2\omega_1)\\
 &+\cos(\omega_4)^2\cos(2\omega_2)\\
A_2^\text{s}(\omega_1, \omega_2, \omega_3, \omega_4) =
                         & \cos(\omega_3)^2\sin(2\omega_1)\\
                         &+\cos(\omega_4)^2\sin(2\omega_2)\\
A_3^\text{c}(\omega_1, \omega_2, \omega_3, \omega_4) =
       & \cos(\omega_3)^2\cos(\omega_4)^2 \cos(2\omega_1)\\
       &+\sin(\omega_3)^2\sin(\omega_4)^2 \cos(2\omega_2)\\
A_3^\text{s}(\omega_1, \omega_2, \omega_3, \omega_4) =
            & \cos(\omega_3)^2\cos(\omega_4)^2 \sin(2\omega_1)\\
            &+\sin(\omega_3)^2\sin(\omega_4)^2 \sin(2\omega_2)\,.
\end{aligned}
\end{equation}

The functions appearing in $Q_2$ are defined as
\begin{equation}
\begin{aligned}
N_2(\omega_1,  \omega_2, \omega_3, \omega_4, \omega_5, \omega_6) &=\\
 t_1 t_2 \cos(2\omega_1) \sin(2\omega_2)& \sin(\omega_3 - \omega_4)\\
           - \sin(2\omega_1)\big(\sin(\omega_2)^2 
           &\sin(\omega_3 - \omega_4 + \omega_5 + \omega_6)\\
          -\cos(\omega_2)^2&\sin(\omega_3-\omega_4-\omega_5-\omega_6)\\
          -r_2&\sin(\omega_3+\omega_4+\omega_5-\omega_6)\\
          +r_1&\sin(\omega_3+\omega_4+\omega_5+\omega_6)\\
          -r_1r_2[\cos(\omega_2)^2&
             \sin(\omega_3 - \omega_4 - \omega_5 + \omega_6)\\
            - \sin(\omega_2)^2&\sin(\omega_3 - \omega_4 + \omega_5 - \omega_6)
          ]\big)\,,
\end{aligned}
\end{equation}
\begin{equation}
\begin{aligned}
    D_2(\omega_2,\omega_3,\omega_5,\omega_4) &=
    [H_2^\text{c}(\omega_2,\omega_3,\omega_5,\omega_4) - 1]^2\\
    &\quad+H_2^\text{s}(\omega_2,\omega_3,\omega_5,\omega_4)^2\,,
\end{aligned}
\end{equation}
\begin{equation}
\begin{aligned}
H_2^\text{c}(\omega_2, \omega_3, \omega_5, \omega_4) &=
   r_1A_4^\text{c}(\omega_2,\omega_3,\omega_4) \\
 &\quad+r_2A_4^\text{c}(\omega_2,\omega_2+\omega_4,\omega_1+\omega_5)\\
 &\quad-r_1r_2 \cos[2(\omega_3+\omega_5+\omega_4)]\\
H_2^\text{s}(\omega_2, \omega_3, \omega_5, \omega_4) &=
   r_1A_4^\text{s}(\omega_2,\omega_3,\omega_4) \\
 &\quad+r_2A_4^\text{s}(\omega_2,\omega_2+\omega_4,\omega_1+\omega_5)\\
 &\quad-r_1r_2 \sin[2(\omega_3+\omega_5+\omega_4)]\,,
\end{aligned}
\end{equation}
and
\begin{equation}
\begin{aligned}
A_4^\text{c}(\omega_2,\omega_3,\omega_4) 
      &=\cos(\omega_2)^2\cos(2\omega_3)+\sin(\omega_2)^2\cos(2\omega_4)\\
A_4^\text{s}(\omega_2,\omega_3,\omega_4) 
      &=\cos(\omega_2)^2\sin(2\omega_3)+\sin(\omega_2)^2\sin(2\omega_4)\,.
\end{aligned}
\end{equation}

\end{document}